\documentclass[11pt]{iopart}

\usepackage{diagrams}
\usepackage{graphicx}
\usepackage{wrapfig}
\usepackage{appendix}
\usepackage{subfigure}

\usepackage{tikz}
\usetikzlibrary{shapes,arrows}

\usepackage{tikzquanto}
\tikzstyle{every loop}=[]

\begin{document}

\newcommand{\me}{\ensuremath{\mathrm{e}}}
\newcommand{\mi}{\ensuremath{\mathrm{i}}}

\newcommand{\denote}[1]{
\[ #1 \]}

\newcommand{\inlinegraphic}[2]{
  \dimendef\grafheight=255\dimendef\grafvshift=254
  \grafheight=#1
  \grafvshift=-0.5\grafheight
  \advance\grafvshift by 0.5ex
  \raisebox{\grafvshift}{\includegraphics[height=\grafheight]{images/#2}\xspace}
}

\newcommand{\inline}[1]{
  \raisebox{0.5ex}{\;#1\;}
}

\newcommand{\greenspider}[1]{%
\begin{tikzpicture}[quanto]
  \spider{green vertex}{spideri}{0,0}
  \node [green angle] at (spideri) {#1};
\end{tikzpicture}
}

\newcommand{\redspider}[1]{%
\begin{tikzpicture}[quanto]
  \spider{red vertex}{spideri}{0,0}
  \node [red angle] at (spideri) {#1};
\end{tikzpicture}
}

\newcommand{\hgate}{%
\begin{tikzpicture}[quanto]
  \node [boundary vertex] (a) at (1.05,-1.0) {};
  \node [hadamard vertex] (c) at (1.05,-2.25) {};
  \node [boundary vertex] (d) at (1.05,-3.525) {};
  \draw [] (a) to (c);
  \draw [] (c) to (d);
\end{tikzpicture}
}

\newcommand{\greenunit}{%
  \begin{tikzpicture}[quanto]
    \node [boundary vertex] (a) at (1.0,-2.0) {};
    \node [green vertex] (b) at (1.0,-1.0) {};
    \draw [] (b) to (a);
  \end{tikzpicture}
}

\newcommand{\redunit}{%
  \begin{tikzpicture}[quanto]
    \node [boundary vertex] (a) at (1.0,-2.0) {};
    \node [red vertex] (b) at (1.0,-1.0) {};
    \draw [] (b) to (a);
  \end{tikzpicture}
}

\newcommand{\greenphase}[1]{%
  \begin{tikzpicture}[quanto]
    \node [green vertex] (c) at (1.175,-2.0) {};
    \node [boundary vertex] (b) at (1.175,-3.0) {};
    \node [boundary vertex] (a) at (1.175,-1.0) {};
    \draw [] (a) to (c);
    \draw [] (c) to (b);
    \node [green angle] at (c) {$#1$};
  \end{tikzpicture}
}

\newcommand{\greenphasesig}[2]{%
  \begin{tikzpicture}[quanto]
    \node [green vertex] (c) at (1.175,-2.0) {};
    \node [boundary vertex] (b) at (1.175,-3.0) {};
    \node [boundary vertex] (a) at (1.175,-1.0) {};
    \draw [] (a) to (c);
    \draw [] (c) to (b);
    \node [green angle] at (c) {$#1,#2$};
  \end{tikzpicture}
}

\newcommand{\redphase}[1]{%
  \begin{tikzpicture}[quanto]
    \node [red vertex] (c) at (1.175,-2.0) {};
    \node [boundary vertex] (b) at (1.175,-3.0) {};
    \node [boundary vertex] (a) at (1.175,-1.0) {};
    \draw [] (a) to (c);
    \draw [] (c) to (b);
    \node [red angle] at (c) {$#1$};
  \end{tikzpicture}
}

\newcommand{\redphasesig}[2]{%
  \begin{tikzpicture}[quanto]
    \node [red vertex] (c) at (1.175,-2.0) {};
    \node [boundary vertex] (b) at (1.175,-3.0) {};
    \node [boundary vertex] (a) at (1.175,-1.0) {};
    \draw [] (a) to (c);
    \draw [] (c) to (b);
    \node [red angle] at (c) {$#1,#2$};
  \end{tikzpicture}
}

\newcommand{\czed}{%
  \begin{tikzpicture}[quanto]
    \node [boundary vertex] (a) at (1.0,-1.0) {};
    \node [boundary vertex] (d) at (3.0,-3.0) {};
    \node [boundary vertex] (c) at (1.0,-3.0) {};
    \node [boundary vertex] (b) at (3.0,-1) {};
    \node [hadamard vertex] (g) at (2.0,-2.0) {};
    \node [green vertex] (f) at (1.0,-2.0) {};
    \node [green vertex] (e) at (3.0,-2.0) {};
    \draw [] (f) to (g);
    \draw [] (e) to (d);
    \draw [] (f) to (c);
    \draw [] (b) to (e);
    \draw [] (a) to (f);
    \draw [] (g) to (e);
  \end{tikzpicture}
}

\newcommand{\meas}{%
  \begin{tikzpicture}[quanto]
    \node [green vertex] (c) at (1.3,-2.25) {};
    \node [boundary vertex] (d) at (1.3,-1.0) {};
    \node [green vertex] (b) at (1.3,-3.5) {};
    \draw [] (c) to (b);
    \draw [] (d) to (c);
    \node [green angle] at (c) {$\pi,\{i\}$};
    \node [green angle] at (b) {$-\alpha$};
  \end{tikzpicture}
}

\newcommand{\ket}[1]{\ensuremath{|#1 \rangle}}
\newcommand{\bra}[1]{\ensuremath{\langle #1|}}

\title{Quantum picturalism for topological cluster-state computing}

\author{Dominic Horsman}
\address{Oxford University Computing Laboratory, Parks Road, Oxford OX1 3QD, UK.\\
Keio University Shonan Fujisawa Campus, Fujisawa, Kanagawa 252-0882, Japan.}
\vspace{10pt}
\begin{indented}
\item[]24 August 2011
\end{indented}

\begin{abstract}
Topological quantum computing is a way of allowing precise quantum computations to run on noisy and imperfect hardware. One implementation uses \emph{surface codes} created by forming defects in a highly-entangled cluster state. Such a method of  computing is a leading candidate for large-scale quantum computing. However, there has been a lack of sufficiently powerful high-level languages to describe computing in this form without resorting to single-qubit operations, which quickly become prohibitively complex as the system size increases. In this paper we apply the category-theoretic work of Abramsky and Coecke to the topological cluster-state model of quantum computing to give a high-level graphical language that enables direct translation between quantum processes and physical patterns of measurement in a computer -- a ``compiler language". We give the equivalence between the graphical and topological information flows, and show the applicable rewrite algebra for this computing model. We show that this gives us a native graphical language for the design and analysis of topological quantum algorithms, and finish by discussing the possibilities for automating this process on a large scale. 

\end{abstract}
\maketitle

\section{Introduction}

Quantum computations running on physically realistic hardware will always be susceptible to error. Components fail, quantum systems decohere, and operations are performed imperfectly. Because of the fragility of quantum data, these errors need to be corrected for any large-scale computation to succeed \cite{shorec}. \emph{Topological quantum computing} is a scheme that allows precise quantum algorithms to be performed using noisy and imperfect physical implementations \cite{kitaev}. Single quantum bits (or \emph{qubits}) are encoded \emph{topologically}, in the global state of a set of qubits, rather than locally definable systems. The required degrees of freedom are produced either by creating \emph{defects} in a lattice or \emph{excitations} from a ground state. This encoding enables the logical and physical qubit levels to be split, enabling physical errors to be corrected. While there are standard patterns of defects or excitations that can be used to produce the action of gates on logical qubits, the reverse operation has been difficult: given a particular pattern, what algorithm does it produce? If we deform a pattern will it still give the same output? Which other patterns  produce the same algorithm? Are there better ways of combining the standard patterns into large-scale algorithms to allow us to optimise for a given set of resources? Up to now, answering such questions has required fine-grained calculation of individual physical states, becoming intractable for even small numbers of qubits.

In this paper we introduce a graphical ``compiler language'' for topological quantum computing, enabling us to translate directly between patterns of defects and logical gate operations at a large scale, without needing to calculate the state of individual qubits. We will use a form of what is known as \emph{quantum picturalism}: a graphical language for defining quantum processes that is intuitively simple and yet supports direct calculation from diagrams alone \cite{monster,bobsamson1,bob1}. Quantum picturalism has had a great deal of success in simplifying and elucidating various aspects of quantum theory, in particular within the field of quantum computation. The successor to Penrose's tensor diagram calculus \cite{penrose}, quantum picturalism emerges when quantum mechanical processes are defined as elements in a \emph{dagger-symmetric monoidal category} \cite{joyal,selinger}, and it is this relation that underwrites the diagrams as a calculus rather than merely a graphical representation. Diagrams that are re-writes of each other according to the given ruleset are mathematically identical; diagrammatic equivalence both entails and is entailed by algebraic equivalence, lending diagram re-writes the status of equations. 

We show in this paper that quantum picturalism also has another property, this time specific to topological computing: that the geometric layout of diagrams exactly matches the physical patterns within a topological algorithm. By re-writing the diagrams we can directly produce different physical layouts for the same algorithm, and conversely any defect pattern can be analysed by converting it directly to a diagram, which can then be re-written into a `normal form', showing the implemented gate sequence. Provided only that the defects are well-formed, this is performed at purely a logical level, requiring no single- (physical) qubit operators.

Topological quantum computing (QC) comes in two distinct yet related forms. Either elementary particles, known as \emph{anyons}, are formed as excited states of an exotic Hamiltonian \cite{anyon,jiannis}, or else defects are defined with respect to highly-entangled lattices of physical qubits in what are called \emph{surface codes} \cite{bravyikitaev,surface}. In both cases, gates are produced by braiding the defects/excitations, thereby changing the topology of the system and hence the encoded information. In this paper we will concentrate for the sake of concreteness on the surface code implementation of topological QC, where qubits are formed by creating holes (the defects) within a regular \emph{cluster state} of entangled qubits. Logical states therefore are comprised of the state of many physical, entangled, qubits. Errors on individual systems can be both detected and corrected, thus protecting the encoded logical qubits. Computation proceeds by sequential measurements of the cluster state, making this a form of error-corrected \emph{measurement-based quantum computing} (MBQC) \cite{MBQC}.

Quantum picturalism has already found significant success in the analysis and construction of measurement-based quantum protocols.  Category-theory based diagrammatics demonstrates the structural connections between MBQC and the circuit model, using the \emph{rewrite rules} of the graphical language to transform MBQC processes to the diagrams associated with the equivalent quantum circuits \cite{ross}. It can also be used to determine the correctness of a given measurement pattern, and whether it can be implemented deterministically or not \cite{rosssimon}. One particularly interesting result states that, although quantum picturalism represents \emph{processes} rather than \emph{states} of individual qubits, there is nevertheless an equivalence between the physical geometry of a given cluster state and the geometry of the diagram representing the process of creating it \cite{rosssimon}. It has previously been suggested that a similar geometrical equivalence is also present in topological cluster-state computing, particularly in 3 dimensions \cite{raussendorf}. When first introduced, attention was drawn to the apparent similarities between the defect geometries and that of the processes represented in quantum picturalism. However, at the time there was not a sufficiently rich pictorial language in which to describe the defects and define this postulated identity. 

 In this paper we use the recently-developed \emph{red/green calculus of observables} \cite{rg,monster} to demonstrate the topological equivalence of the defect lines in 3D cluster state computing and the equivalent category-theoretic flow diagrams. We first introduce the calculus (\S \ref{QP}), then show how topological QC can be cast into red/green form in \S\ref{TQC}. We show how the rewrite rules of the calculus define a single data flow line from the multiple qubits comprising the defect and correlation surface, and give simple graphical descriptions of logical operators, including stabilizers, around them. We give the graphical representations of braiding between qubits, and the roles played in the computation by the primal and dual lattices. We finish by discussing the use of the calculus as an automated design and analysis tool for topolgical algorithms. In this paper we will concentrate on the graphical representation, and not require knowledge of the underlying mathematics of monoidal categories and Frobenius algebras. The reader is reminded, though, that diagrams are formed from a fully rigorous calculus, and that each picture represents in itself a well-defined equation or mathematical expression.    

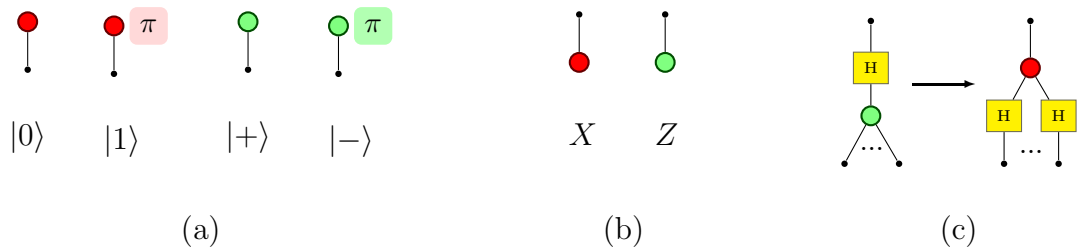
\begin{figure}[tb]
   \centering
	\hspace{-4cm} \scalebox{1}{ \begin{minipage}[c]{1.0\linewidth}
  \centering
  \[
 \begin{array}{ccccc}

 \begin{array}{c}
        \begin{tikzpicture}[quanto]
          \node [red vertex] (b) at (1.175,-2.25) {}; \node [boundary
           vertex] (a) at (1.175,-3.525) {}; \draw [] (b) to (a);
        \end{tikzpicture} \\ {} \\
        \ket{0}
        
  \end{array}
      \quad
   \begin{array}{l}
         \begin{tikzpicture}[quanto]
           \node [red vertex] (b) at (1.175,-2.25) {}; \node [boundary
           vertex] (a) at (1.175,-3.525) {}; \draw [] (b) to (a);
           \node [red angle] at (b) {$\pi$};
        \end{tikzpicture}\\{}\\\ket{1}\\
\end{array}           
\quad
 \begin{array}{c}
        \begin{tikzpicture}[quanto]
          \node [green vertex] (b) at (1.175,-2.25) {}; \node [boundary
           vertex] (a) at (1.175,-3.525) {}; \draw [] (b) to (a);
        \end{tikzpicture} \\ {} \\
        \ket{+}
        
  \end{array}
      \quad
   \begin{array}{l}
         \begin{tikzpicture}[quanto]
           \node [green vertex] (b) at (1.175,-2.25) {}; \node [boundary
           vertex] (a) at (1.175,-3.525) {}; \draw [] (b) to (a);
           \node [green angle] at (b) {$\pi$};
        \end{tikzpicture}\\{}\\\ket{-}\\
\end{array}  

& \qquad
\quad &

\begin{array}{l}
         \begin{tikzpicture}[quanto]
           \node [boundary vertex] (b) at (1.175,-2.25) {}; \node [red
           vertex] (a) at (1.175,-3.525) {}; \draw [] (b) to (a);
        \end{tikzpicture}\\{}\\X\\
\end{array}   
\quad
\begin{array}{l}
         \begin{tikzpicture}[quanto]
           \node [boundary vertex] (b) at (1.175,-2.25) {}; \node [green
           vertex] (a) at (1.175,-3.525) {}; \draw [] (b) to (a);
        \end{tikzpicture}\\{}\\Z\\
\end{array}  

& \qquad
\quad & 


\begin{tikzpicture}[quanto]
\node [boundary vertex] (b) at (1.05,-4.8) {};
\node [boundary vertex] (c) at (2.5,-4.8) {};
\node [green vertex] (e) at (1.75,-3.525) {};
\node [hadamard vertex] (d) at (1.75,-2.25) {};
\node [boundary vertex] (a) at (1.75,-1.0) {};
\node [ellipses] (f) at ($ 0.5*(b) + 0.5*(c) + (0,0.4)$) {} ;
\draw [] (e) to (c);
\draw [] (a) to (d);
\draw [] (e) to (b);
\draw [] (d) to (e);
\end{tikzpicture}
\rTo
\begin{tikzpicture}[quanto]
\node [boundary vertex] (a) at (1.75,-1.0) {};
\node [hadamard vertex] (f) at (1.05,-3.525) {};
\node [boundary vertex] (b) at (1.05,-4.8) {};
\node [boundary vertex] (c) at (2.5,-4.8) {};
\node [red vertex] (d) at (1.75,-2.25) {};
\node [hadamard vertex] (e) at (2.5,-3.525) {};
\node [ellipses] (g) at ($ 0.5*(b) + 0.5*(c) + (0,0.33)$) {} ;
\draw [] (f) to (b);
\draw [] (a) to (d);
\draw [] (d) to (f);
\draw [] (d) to (e);
\draw [] (e) to (c);
\end{tikzpicture}
\\{}\\
\mathrm{(a)} & \qquad \quad &  \mathrm{(b)} & \qquad \quad & \mathrm{(c)}
\end{array}
\]
\end{minipage}}
   \caption{The fundamental elements in the red/green calculus: a) state preparation; b) measurement; c) the Hadamard box and branching.} \label{fund_elem}
\end{figure}

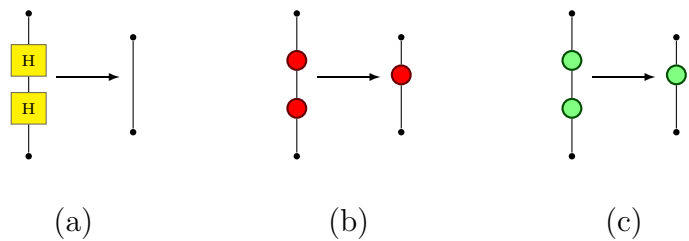
\begin{figure}[t]
   \centering
    \scalebox{1}{ \begin{minipage}[c]{1.0\linewidth}
  \centering
  \[
 \begin{array}{ccccc}

\begin{tikzpicture}[quanto]
\node [boundary vertex] (a) at (1.05,-1.0) {};
\node [hadamard vertex] (c) at (1.05,-2.25) {};
\node [hadamard vertex] (d) at (1.05,-3.525) {};
\node [boundary vertex] (b) at (1.05,-4.8) {};
\draw [] (d) to (b);
\draw [] (a) to (c);
\draw [] (c) to (d);
\end{tikzpicture}
\rTo
\begin{tikzpicture}[quanto]
\node [boundary vertex] (c) at (1.05,-3.525) {};
\node [boundary vertex] (a) at (1.05,-1.0) {};
\draw [] (a) to (c);
\end{tikzpicture}

& \qquad
\quad &

\begin{tikzpicture}[quanto]
\node [boundary vertex] (a) at (1.05,-1.0) {};
\node [red vertex] (c) at (1.05,-2.25) {};
\node [red vertex] (d) at (1.05,-3.525) {};
\node [boundary vertex] (b) at (1.05,-4.8) {};
\draw [] (d) to (b);
\draw [] (a) to (c);
\draw [] (c) to (d);
\end{tikzpicture}
\rTo
\begin{tikzpicture}[quanto]
\node [boundary vertex] (c) at (1.05,-3.525) {};
\node [red vertex] (b) at (1.05,-2) {};
\node [boundary vertex] (a) at (1.05,-1.0) {};
\draw [] (a) to (b);
\draw [] (b) to (c);
\end{tikzpicture}

& \qquad
\quad & 


\begin{tikzpicture}[quanto]
\node [boundary vertex] (a) at (1.05,-1.0) {};
\node [green vertex] (c) at (1.05,-2.25) {};
\node [green vertex] (d) at (1.05,-3.525) {};
\node [boundary vertex] (b) at (1.05,-4.8) {};
\draw [] (d) to (b);
\draw [] (a) to (c);
\draw [] (c) to (d);
\end{tikzpicture}
\rTo
\begin{tikzpicture}[quanto]
\node [boundary vertex] (c) at (1.05,-3.525) {};
\node [green vertex] (b) at (1.05,-2) {};
\node [boundary vertex] (a) at (1.05,-1.0) {};
\draw [] (a) to (b);
\draw [] (b) to (c);
\end{tikzpicture}
\\{}\\
\mathrm{(a)} & \qquad \quad &  \mathrm{(b)} & \qquad \quad & \mathrm{(c)}
\end{array}
\]
\end{minipage}}
   \caption{Composition rules for the three elements in the calculus} \label{comp}
\end{figure}

\section{Quantum picturalism}\label{QP}
We outline here a simplified form of the red/green calculus of \cite{rg,monster}. The diagrams represent \emph{processes} rather than \emph{qubits}: a straight line is the identity operation, and the other elements more complex processes. They therefore describe the \emph{dependencies} of the processes within quantum protocols: processes joined by lines depend on each other, with the output of one becoming the input of the next. All such processes can be \emph{composed}: a single input and output given for them (we will see that this is the basis for many of the rules). Formally, for a process $P_1$ that maps $i$ inputs to $j$ outputs, and a second process $P_2$ mapping $j$ inputs to $k$ outputs, the composition of the two $P_1\circ P_2$ maps $i$ inputs to $k$ outputs. We adopt the convention of time running from top to bottom. Scalar normalisation is neglected throughout.\\
\indent The fundamental elements in the formalism are the $X$ and $Z$ observables (this makes the red/green calculus a natural choice for use with cluster state computing, which is usually described using the $X$ and $Z$ stabilizers).  Figure \ref{fund_elem}(a) shows state preparation in the two bases. Conversely, figure \ref{fund_elem}(b) shows measurement in these bases. Measurement in quantum picturalism is always ``post-selected'' on the result of the measurement, and conventionally taken to represent the `+1' outcome. The final element is the Hadamard box, shown in figure \ref{fund_elem}(c), which converts between them. Defining the Hadamard box in this form also introduces the notion that  given process can have more that one input or output (that is, many straight lines either emerging or entering). Again, these demonstrate the way different processes depend on each other, and enable us to compose their effects.

\begin{figure}[t]
   \centering
     \scalebox{1}{ \begin{minipage}[c]{1.0\linewidth}
  \centering
  \[
\begin{tikzpicture}[quanto]
  \upspider{green vertex}{spideri}{0,0}
  \downspider{green vertex}{spiderii}{0,-1.5}
  \draw [] (spideri) to (spiderii) ;
  \node [green angle] at (spideri) {$\alpha $};
  \node [green angle] at (spiderii) {$\beta $};
\end{tikzpicture}
\rTo
\begin{tikzpicture}[quanto]
\spider{green vertex}{a}{0,0}
\node [green angle] at (a) {$\alpha + \beta $};
\end{tikzpicture}
\]
\end{minipage}}
   \caption{The fundamental combination rule: the `spider' rule} \label{combin1}
\end{figure}
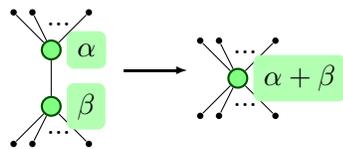

\begin{figure}[t]
   \centering
   \hspace{-4cm}   \scalebox{1}{ \begin{minipage}[c]{1.0\linewidth}
  \centering
  \[
 \begin{array}{ccccc}

\begin{tikzpicture}[quanto]
\node [boundary vertex] (b) at (1.05,-4.8) {};
\node [boundary vertex] (c) at (2.5,-4.8) {};
\node [green vertex] (e) at (1.75,-3.525) {};
\node [red vertex] (d) at (1.75,-2.25) {};
\node [boundary vertex] (a) at (1.75,-1.0) {};
\node [ellipses] (f) at ($ 0.5*(b) + 0.5*(c) + (0,0.4)$) {} ;
\draw [] (e) to (c);
\draw [] (a) to (d);
\draw [] (e) to (b);
\draw [] (d) to (e);
\end{tikzpicture}
\rTo
\begin{tikzpicture}[quanto]
\node [boundary vertex] (a) at (1.75,-1.0) {};
\node [red vertex] (f) at (1.05,-3.525) {};
\node [boundary vertex] (b) at (1.05,-4.8) {};
\node [boundary vertex] (c) at (2.5,-4.8) {};
\node [green vertex] (d) at (1.75,-2.25) {};
\node [red vertex] (e) at (2.5,-3.525) {};
\node [ellipses] (g) at ($ 0.5*(b) + 0.5*(c) + (0,0.33)$) {} ;
\draw [] (f) to (b);
\draw [] (a) to (d);
\draw [] (d) to (f);
\draw [] (d) to (e);
\draw [] (e) to (c);
\end{tikzpicture}

& \qquad
\quad &

\begin{tikzpicture}[quanto]
\node [boundary vertex] (b) at (1.05,-4.8) {};
\node [boundary vertex] (c) at (2.5,-4.8) {};
\node [green vertex] (e) at (1.75,-3.525) {};
\node [red vertex] (d) at (1.75,-2.25) {};
	\node [red angle] at (d) {$\pi$};
\node [boundary vertex] (a) at (1.75,-1.0) {};
\node [ellipses] (f) at ($ 0.5*(b) + 0.5*(c) + (0,0.4)$) {} ;
\draw [] (e) to (c);
\draw [] (a) to (d);
\draw [] (e) to (b);
\draw [] (d) to (e);
\end{tikzpicture}
\rTo
\begin{tikzpicture}[quanto]
\node [boundary vertex] (a) at (1.75,-1.0) {};
\node [red vertex] (f) at (1.05,-3.525) {};
	\node [red angle] at (f) {$\pi$};
\node [boundary vertex] (b) at (1.05,-4.8) {};
\node [boundary vertex] (c) at (2.5,-4.8) {};
\node [green vertex] (d) at (1.75,-2.25) {};
\node [red vertex] (e) at (2.5,-3.525) {};
	\node [red angle] at (e) {$\pi$};
\node [ellipses] (g) at ($ 0.5*(b) + 0.5*(c) + (0,0.33)$) {} ;
\draw [] (f) to (b);
\draw [] (a) to (d);
\draw [] (d) to (f);
\draw [] (d) to (e);
\draw [] (e) to (c);
\end{tikzpicture}

& \qquad
\quad &

\begin{tikzpicture}[quanto]
\node [boundary vertex] (b) at (1.05,-4.8) {};
\node [boundary vertex] (c) at (2.5,-4.8) {};
\node [green vertex] (e) at (1.75,-3.525) {};
\node [red vertex] (d) at (1.75,-2.25) {};
\node [ellipses] (f) at ($ 0.5*(b) + 0.5*(c) + (0,0.4)$) {} ;
\draw [] (e) to (c);
\draw [] (e) to (b);
\draw [] (d) to (e);
\end{tikzpicture}
\rTo
\begin{tikzpicture}[quanto]
\node [red vertex] (f) at (1.05,-3.525) {};
\node [boundary vertex] (b) at (1.05,-4.8) {};
\node [boundary vertex] (c) at (2.5,-4.8) {};
\node [red vertex] (e) at (2.5,-3.525) {};
\node [ellipses] (g) at ($ 0.5*(b) + 0.5*(c) + (0,0.33)$) {} ;
\draw [] (f) to (b);
\draw [] (e) to (c);
\end{tikzpicture}

\\{}\\
 \mathrm{(a)} & \qquad \quad  &  \mathrm{(b)}  & \qquad \quad & \mathrm{(c)} 
\end{array}
\]
\end{minipage}}
   \caption{The `copying' combination rules} \label{combin}
\end{figure}
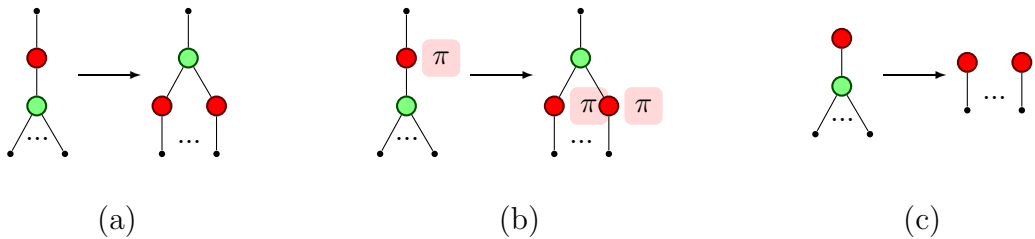

We can now give the first example of composition rules for these elements. The diagrams in figure \ref{comp} can all be read as ``what happens to the output when the same process is repeated twice?". Figure \ref{comp}(a) is immediately obvious: applying a Hadamard operator twice is the same as applying the identity operator. The remaining composition rules should become evident as we continue to introduce the scheme.

\indent The most important composition rule for these elements is the \emph{spider rule}, figure \ref{combin1}. This takes $m$ inputs to $n$ outputs dependent on the phases of the intervening nodes. The diagram here can be described by $\ket{0}^{\otimes n}\bra{0}^{\otimes m} + \me^{\mi (\alpha + \beta)}\ket{1}^{\otimes n}\bra{1}^{\otimes m} $. When red nodes approach a green spider, we can see that they are `copied', figure \ref{combin}. In each case the green node is $\delta_{green} = \ket{00}\bra{0} + \ket{11}\bra{1}$, which takes $\ket{0} \rightarrow \ket{00}$, $\ket{1} \rightarrow \ket{11}$. All these rules hold with green and red colours reversed, and the $\{\ket{0},\ket{1}\}$ basis swapped for the $\{\ket{+},\ket{-}\}$. These `copying' operations can be read intuitively as showing what operations on the $n$ outputs is equivalent to the single operation shown on the single input. Recalling that vertices represent processes rather than qubits will avoid any potential confusion with the no-cloning theorem.

\indent As a first application of the calculus, we can give the representations of some familiar quantum gates. Figure \ref{gates} shows the controlled-NOT (CNOT) and controlled-Z (CZ) gates. An unusual and useful feature of the red/green calculus is that lines between qubits are non-directional. As an example, we can `read' the CNOT in figure \ref{gates}(a). The green node copies the input on the left-hand qubit rail in the computational basis, as $\delta_{green}$ above, and the red copies the $\{\ket{+},\ket{-}\}$ basis. If we define the left and right vertical rails as 1 and 3, and the centre as 2, then we have the action of the diagram as 
\begin{eqnarray*}
\Big ( (\ket{00}_{12}\bra{0}_1 + \ket{11}_{12}\bra{1}_1)  \otimes (\ket{+}_3\bra{++}_{23} + \ket{-}_3\bra{--}_{23} ) \Big ) \ket{\psi}_{13}\ \ \ \ \ \ \ \ \ \ \ \ \ \  & {} &\\
 = \Big (\ket{0}\bra{0}_1 \otimes (\ket{+}\bra{+}_3 + \ket{-}\bra{-}_3) + \ket{1}\bra{1}_1 \otimes (\ket{+}\bra{+}_3 - \ket{-}\bra{-}_3)\Big ) \ket{\psi}_{13} & {} &\\
= \Big ( \ket{0}\bra{0}_1 \otimes (\ket{0}\bra{0}_3 + \ket{1}\bra{1}_3) \ + \ \ket{1}\bra{1}_1 \otimes (\ket{0}\bra{1}_3 + \ket{1}\bra{0}_3) \Big )  \ket{\psi}_{13} \ \ \  \ \ & {} &
 \end{eqnarray*}

\noindent where $ \ket{\psi}_{13}$ has support only on the Hilbert space of qubits 1 and 3. We can see that this is the action of the CNOT gate. From this we can derive the representation of the CZ gate, using the gate equivalence CNOT = H$_2$CNOT H$_2$. Figure \ref{gates}(b) shows the representation of this, and the final `normal form' after applying the re-write rule of figure  \ref{fund_elem}(c).

\section{Topological cluster state computing}\label{TQC}
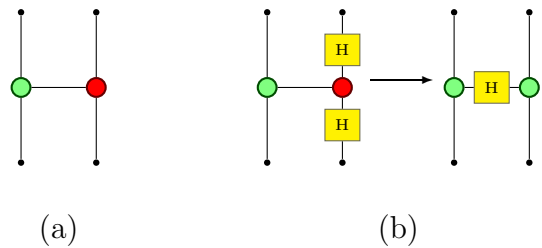
\begin{figure}[t]
   \centering
       \scalebox{1}{ \begin{minipage}[c]{1.0\linewidth}
  \centering
  \[
 \begin{array}{ccc}

\begin{tikzpicture}[quanto]
\node [boundary vertex] (a) at (0.0,-1.0) {};
\node [boundary vertex] (w) at (2.0,-1.0) {};
\node [green vertex] (spider2) at (0.0,1.0) {};
\node [red vertex] (j) at (2.0,1.0) {};
\node [boundary vertex] (b) at (0.0,3.0) {};
\node [boundary vertex] (c) at (2.0,3.0) {};
\draw [] (spider2) to (a);
\draw [] (c) to (j);
\draw [] (w) to (j);
\draw [] (b) to (spider2);
\draw [] (j) to (spider2);
\end{tikzpicture}

& \qquad
\quad &
\begin{tikzpicture}[quanto]
\node [boundary vertex] (a) at (0.0,-1.0) {};
\node [boundary vertex] (w) at (2.0,-1.0) {};
\node [green vertex] (spider2) at (0.0,1.0) {};
\node [hadamard vertex] (g) at (2.0,2.0) {};
\node [hadamard vertex] (h) at (2.0,0) {};
\node [red vertex] (j) at (2.0,1.0) {};
\node [boundary vertex] (b) at (0.0,3.0) {};
\node [boundary vertex] (c) at (2.0,3.0) {};
\draw [] (spider2) to (a);
\draw [] (c) to (g);
\draw [] (g) to (j);
\draw [] (w) to (h);
\draw [] (h) to (j);
\draw [] (b) to (spider2);
\draw [] (j) to (spider2);
\end{tikzpicture}
\rTo
\begin{tikzpicture}[quanto]
\node [boundary vertex] (a) at (0.0,-1.0) {};
\node [boundary vertex] (w) at (2.0,-1.0) {};
\node [green vertex] (spider2) at (0.0,1.0) {};
\node [hadamard vertex] (g) at (1.0,1.0) {};
\node [green vertex] (j) at (2.0,1.0) {};
\node [boundary vertex] (b) at (0.0,3.0) {};
\node [boundary vertex] (c) at (2.0,3.0) {};
\draw [] (spider2) to (a);
\draw [] (c) to (j);
\draw [] (w) to (j);
\draw [] (b) to (spider2);
\draw [] (j) to (g);
\draw [] (g) to (spider2);
\end{tikzpicture}
\\{}\\
 \mathrm{(a)} & \qquad \quad  &  \mathrm{(b)}  \end{array}
\]
\end{minipage}}
   \caption{Quantum gates in the calculus: a) CNOT; b) CZ} \label{gates}
\end{figure}
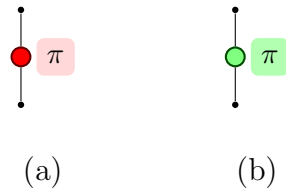
\begin{figure}[t]
   \centering
       \scalebox{1}{ \begin{minipage}[c]{1.0\linewidth}
  \centering

  \[
 \begin{array}{ccc}

\begin{tikzpicture}[quanto]
        \node [red vertex] (c) at (1.175,-2.25) {}; \node [boundary
        vertex] (b) at (1.175,-3.525) {}; \node [boundary vertex] (a)
        at (1.175,-1.0) {}; \draw [] (a) to (c); \draw [] (c) to (b);
        \node [red angle] at (c) {$\pi$};
      \end{tikzpicture}

& \qquad
\quad &

\begin{tikzpicture}[quanto]
        \node [green vertex] (c) at (1.175,-2.25) {}; \node [boundary
        vertex] (b) at (1.175,-3.525) {}; \node [boundary vertex] (a)
        at (1.175,-1.0) {}; \draw [] (a) to (c); \draw [] (c) to (b);
        \node [green angle] at (c) {$\pi$};
      \end{tikzpicture}
\\{}\\
 \mathrm{(a)} & \qquad \quad  &  \mathrm{(b)}  \end{array}
\]
\end{minipage}}
   \caption{Operators in the calculus: a) $X$ and b) $Z$.} \label{stabs}
\end{figure}

The cluster state computing scheme of \cite{raussendorf} combines the principles of MBQC with topological error correction. It is a 3D version of the \emph{surface code} \cite{surface,austin2D}:  the two dimensions of the cross-section of the lattice encode the qubit data, and the depth of the lattice is the `simulated time' dimension \cite{austin3D}. The cluster state is created by initialising all qubits in the $\ket{+}=\ket{0}+\ket{1}$ state, then performing CZ operations between nearest neighbours, thus determining the geometry. Defects are formed and braided within a 3D cluster state by measuring qubits within the defect boundaries in the $Z$ basis (thus removing them from the lattice). Computation proceeds with the remaining qubits in the next cross-section being measured out in the $X$ basis at the next timestep. Note that, while it is common to talk about the \emph{defects} as logical qubits, they do not in themselves contain any information (being physical qubits removed from the lattice). Rather, the qubits surrounding the defect support the \emph{correlation surface}, which contains the qubit data.

These surface codes have several desirable properties for using them in quantum computing implementations. They can correct the highest proportion of errors in hardware (detection errors, decoherence, etc) of any well-defined implementation -- over 1\% error can be tolerated in each component in the computer \cite{onepc}. The codes also only require nearest neighbouring quantum components to be coupled to form the cluster lattice, removing any need for long-range entanglement. Several architectural designs are in development for both the 2D and 3D topological codes \cite{simon,rod}, and recent experimental advances bring a scalable cluster-state computer closer to reality \cite{exec}.

\subsection{Stabilizers}
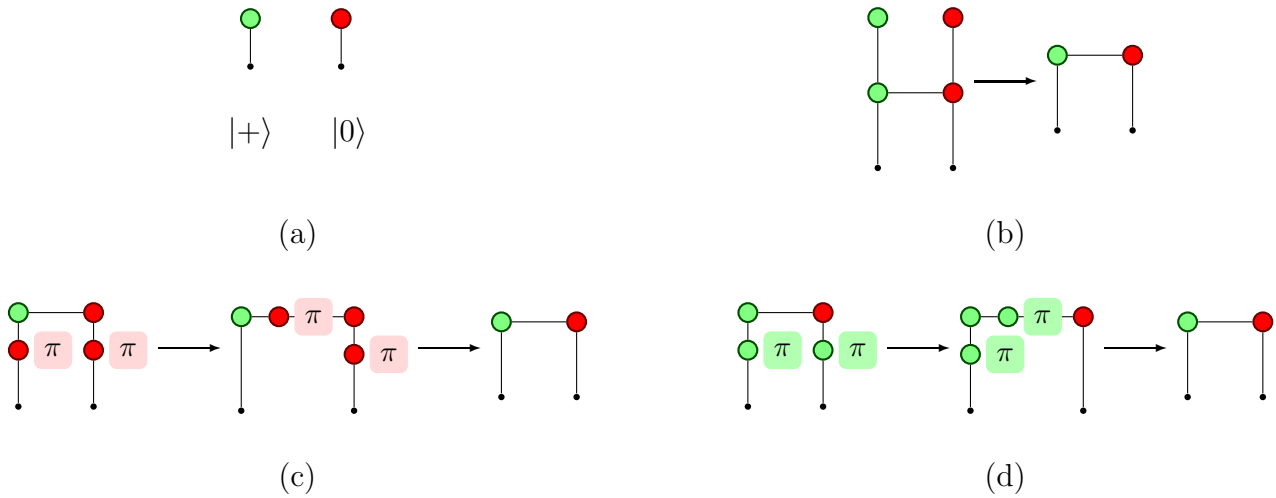
\begin{figure}[t]
   \centering
   \hspace{-4cm}   \scalebox{1}{ \hspace{-2.75cm}   \begin{minipage}[c]{1.0\linewidth}
  \centering
    \[
 \begin{array}{ccc}

 \begin{array}{c}
        \begin{tikzpicture}[quanto]
          \node [green vertex] (b) at (1.175,-2.25) {}; \node [boundary
           vertex] (a) at (1.175,-3.525) {}; \draw [] (b) to (a);
        \end{tikzpicture} \\ {} \\
        \ket{+}
        
  \end{array}
      \quad
   \begin{array}{l}
         \begin{tikzpicture}[quanto]
           \node [red vertex] (b) at (1.175,-2.25) {}; \node [boundary
           vertex] (a) at (1.175,-3.525) {}; \draw [] (b) to (a);
        \end{tikzpicture}\\{}\\ \ket{0}\\
\end{array}   

& \qquad
\quad &

\begin{tikzpicture}[quanto]
\node [boundary vertex] (a) at (0.0,-1.0) {};
\node [boundary vertex] (w) at (2.0,-1.0) {};
\node [green vertex] (spider2) at (0.0,1.0) {};
\node [red vertex] (j) at (2.0,1.0) {};
\node [green vertex] (b) at (0.0,3.0) {};
\node [red vertex] (c) at (2.0,3.0) {};
\draw [] (spider2) to (a);
\draw [] (c) to (j);
\draw [] (w) to (j);
\draw [] (b) to (spider2);
\draw [] (j) to (spider2);
\end{tikzpicture}
\rTo
\begin{tikzpicture}[quanto]
\node [boundary vertex] (a) at (0.0,-1.0) {};
\node [boundary vertex] (w) at (2.0,-1.0) {};
\node [green vertex] (spider2) at (0.0,1.0) {};
\node [red vertex] (j) at (2.0,1.0) {};
\draw [] (spider2) to (a);
\draw [] (w) to (j);
\draw [] (j) to (spider2);
\end{tikzpicture}
\\{}\\
 \mathrm{(a)} & \qquad \quad  &  \mathrm{(b)} \\{}\\


\begin{tikzpicture}[quanto]
\node [boundary vertex] (a) at (0.0,-0.5) {};
\node [boundary vertex] (w) at (2.0,-0.5) {};
\node [green vertex] (spider2) at (0.0,2.0) {};
\node [red vertex] (j) at (2.0,2.0) {};
\node [red vertex] (b) at (0.0,1.0) {};
\node [red vertex] (c) at (2.0,1.0) {};
\draw [] (spider2) to (b);
\draw [] (b) to (a);
\draw [] (w) to (c);
\draw [] (c) to (j);
\draw [] (j) to (spider2);
\node [red angle] at (b) {$\pi$};
\node [red angle] at (c) {$\pi$};
\end{tikzpicture}
\rTo
\begin{tikzpicture}[quanto]
\node [boundary vertex] (a) at (0.0,-0.5) {};
\node [boundary vertex] (w) at (3.0,-0.5) {};
\node [green vertex] (spider2) at (0.0,2.0) {};
\node [red vertex] (j) at (3.0,2.0) {};
\node [red vertex] (b) at (1.0,2.0) {};
\node [red vertex] (c) at (3.0,1.0) {};
\draw [] (spider2) to (a);
\draw [] (w) to (c);
\draw [] (c) to (j);
\draw [] (j) to (b);
\draw [] (b) to (spider2);
\node [red angle] at (b) {$\pi$};
\node [red angle] at (c) {$\pi$};
\end{tikzpicture}
\rTo
\begin{tikzpicture}[quanto]
\node [boundary vertex] (a) at (0.0,-1.0) {};
\node [boundary vertex] (w) at (2.0,-1.0) {};
\node [green vertex] (spider2) at (0.0,1.0) {};
\node [red vertex] (j) at (2.0,1.0) {};
\draw [] (spider2) to (a);
\draw [] (w) to (j);
\draw [] (j) to (spider2);
\end{tikzpicture}

& \qquad
\quad & 

\begin{tikzpicture}[quanto]
\node [boundary vertex] (a) at (0.0,-0.5) {};
\node [boundary vertex] (w) at (2.0,-0.5) {};
\node [green vertex] (spider2) at (0.0,2.0) {};
\node [red vertex] (j) at (2.0,2.0) {};
\node [green vertex] (b) at (0.0,1.0) {};
\node [green vertex] (c) at (2.0,1.0) {};
\draw [] (spider2) to (b);
\draw [] (b) to (a);
\draw [] (w) to (c);
\draw [] (c) to (j);
\draw [] (j) to (spider2);
\node [green angle] at (b) {$\pi$};
\node [green angle] at (c) {$\pi$};
\end{tikzpicture}
\rTo
\begin{tikzpicture}[quanto]
\node [boundary vertex] (a) at (0.0,-0.5) {};
\node [boundary vertex] (w) at (3.0,-0.5) {};
\node [green vertex] (spider2) at (0.0,2.0) {};
\node [red vertex] (j) at (3.0,2.0) {};
\node [green vertex] (b) at (1.0,2.0) {};
\node [green vertex] (c) at (0.0,1.0) {};
\draw [] (a) to (c);
\draw [] (c) to (spider2);
\draw [] (j) to (b);
\draw [] (b) to (spider2);
\draw [] (w) to (j);
\node [green angle] at (b) {$\pi$};
\node [green angle] at (c) {$\pi$};
\end{tikzpicture}
\rTo
\begin{tikzpicture}[quanto]
\node [boundary vertex] (a) at (0.0,-1.0) {};
\node [boundary vertex] (w) at (2.0,-1.0) {};
\node [green vertex] (spider2) at (0.0,1.0) {};
\node [red vertex] (j) at (2.0,1.0) {};
\draw [] (spider2) to (a);
\draw [] (w) to (j);
\draw [] (j) to (spider2);
\end{tikzpicture}
\\{}\\
 \mathrm{(c)} & \qquad \quad  & \mathrm{(d)}
\end{array}
\]
\end{minipage}}
   \caption{Stabilizers for the state $\ket{00} + \ket{11}$: a) and b), constructing the state; c) $XX$; d) $ZZ$.} \label{stabeg}
\end{figure}

\indent The natural (non-graphical) mathematical description for cluster state computing is in terms of the stabilizers of the system \cite{gottesman}. These are operators of which the system is in an eigenstate. As a first application of the graphical formalism to cluster state computing, we will show how these are represented in the red/green calculus. Figure \ref{stabs} gives the $X$ and $Z$ operators within the calculus. We can see that there is a close relationship between the stabilizer operators and the red and green elements of the calculus; however, the exact nature is slightly subtle. To see this, we will work through the example of figure \ref{stabeg} in detail, giving the stabilizers for the Bell state $\ket{00} + \ket{11}$.

First we prepare the state by performing a CNOT on the state $\ket{+}\ket{0}$. In figure \ref{stabeg}(a) we prepare the two states. Figure \ref{stabeg}(b) shows the application of the CNOT after the state preparation (remember that time runs from top to bottom), and then applications of the composition rules of figures \ref{comp}(a) and \ref{comp}(b). Then we look at the operators $XX$ (figure \ref{stabeg}(c)) and $ZZ$ (figure \ref{stabeg}(d)). These are stabilizers of the system if we can rewrite the state after their addition as the same state as before. The rewrite processes are shown in the figures. The first re-write is an application of the rules given by figure \ref{combin}(b) and its colour-inverse, and the second is an application of the spider rule. This is crucial for intuitively being able to `read off' the stabilizers for a (cluster) state: operators need to be grouped in pairs with a node of the same colour with no phase.

\subsection{Cluster state structure}
\indent  We can now look at the structure of cluster states within the pictorial language. Two important prior results lay the groundwork. Firstly, that any cluster state can be represented pictorially by a geometrically identical diagram \cite{rosssimon}. Secondly, that this diagram is two-colourable with red and green vertices \cite{rosssimon}. Figure \ref{clusterstate} demonstrates these properties. The two colourations given are equivalent; the choice of which to use is a notational decision. Note that as we have a two-dimensional physical cluster, we are representing process time as going ``into the page'', using short diagonal lines.

\begin{figure}[t]
  \centering
    \scalebox{1}{ \begin{minipage}[c]{1.0\linewidth}
  \centering

  \[
\begin{tikzpicture}[quanto]
\node [black vertex] (centre) at (1.5,1.5) {};
\node [black vertex] (a) at (1.5,0.0) {};
\node [black vertex] (b) at (1.5,3.0) {};
\node [black vertex] (c) at (0.0,1.5) {};
\node [black vertex] (d) at (3.0,1.5) {};
\draw [] (a) to (centre);
\draw [] (b) to (centre);
\draw [] (c) to (centre);
\draw [] (d) to (centre);
\end{tikzpicture}
\rTo
\begin{tikzpicture}[quanto]
\node [red vertex] (centre) at (1.5,1.5) {};
\node [green vertex] (a) at (1.5,0.0) {};
\node [green vertex] (b) at (1.5,3.0) {};
\node [green vertex] (c) at (0.0,1.5) {};
\node [green vertex] (d) at (3.0,1.5) {};
\draw [] (a) to (centre);
\draw [] (b) to (centre);
\draw [] (c) to (centre);
\draw [] (d) to (centre);
\node [boundary vertex] (ba) at (1.0,-0.5) {};
\draw [] (a) to (ba);
\node [boundary vertex] (bb) at (1.0,2.5) {};
\draw [] (b) to (bb);
\node [boundary vertex] (bc) at (-0.5,1.0) {};
\draw [] (c) to (bc);
\node [boundary vertex] (bd) at (2.5,1.0) {};
\draw [] (d) to (bd);
\node [hadamard vertex] (h) at (0.75,0.75) {};
\node [boundary vertex] (bh) at (0.25,0.25) {};
\draw [] (centre) to (h);
\draw [] (h) to (bh);
\end{tikzpicture}
\rTo
\begin{tikzpicture}[quanto]
\node [green vertex] (centre) at (2,2) {};
\node [red vertex] (a) at (2,0.0) {};
\node [red vertex] (b) at (2,4) {};
\node [red vertex] (c) at (0.0,2) {};
\node [red vertex] (d) at (4,2) {};
\draw [] (a) to (centre);
\draw [] (b) to (centre);
\draw [] (c) to (centre);
\draw [] (d) to (centre);
\node [hadamard vertex] (ba) at (1.25,-0.75) {};
\node [boundary vertex] (bba) at (0.75,-1.25) {};
\draw [] (ba) to (bba);
\draw [] (a) to (ba);
\node [hadamard vertex] (bb) at (1.25,3.25) {};
\node [boundary vertex] (bbb) at (0.75,2.75) {};
\draw [] (b) to (bb);
\draw [] (bb) to (bbb);
\node [hadamard vertex] (bc) at (-0.75,1.25) {};
\node [boundary vertex] (bbc) at (-1.25,0.75) {};
\draw [] (bc) to (bbc);
\draw [] (c) to (bc);
\node [hadamard vertex] (bd) at (3.25,1.25) {};
\node [boundary vertex] (bbd) at (2.75,0.75) {};
\draw [] (bd) to (bbd);
\draw [] (d) to (bd);
\node [boundary vertex] (bh) at (1,1) {};
\draw [] (centre) to (bh);
\end{tikzpicture}
\]
\end{minipage}}
    \caption{A 2D cluster state written as a two-coloured diagram.}\label{clusterstate}
 \end{figure}
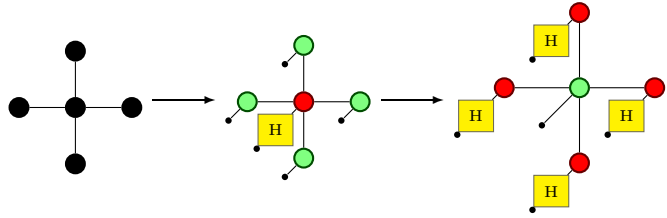
   \begin{figure}[t]
   \centering
  \hspace{-4cm}  \scalebox{0.85}{ \hspace{-7cm}   \begin{minipage}[c]{1.0\linewidth}
 \centering
    \[
 \begin{tikzpicture}
	\node [green vertex] (0) at (9, 3) {};
	\node [black vertex] (1) at (-5.5, 2.5) {};
	\node [boundary vertex] (2) at (8.5, 2.5) {};
	
	\node [green vertex] (3) at (2.75, 2.25) {};
	\node [red vertex] (4) at (2.75, 2.25) {};
	\node [green vertex] (5) at (2.75, 2.25) {};
	
	\node [red vertex] (6) at (4.25, 2.25) {};
	\node [green vertex] (7) at (4.25, 2.25) {};
	\node [red vertex] (8) at (4.25, 2.25) {};
	\node [red vertex] (9) at (5.75, 2.25) {};
	\node [green vertex] (10) at (5.75, 2.25) {};
	\node [black vertex] (11) at (-8, 1.75) {};
	\node [black vertex] (12) at (-6.5, 1.75) {};
	\node [black vertex] (13) at (-5, 1.75) {};
	\node [boundary vertex] (14) at (2.25, 1.75) {};
	\node [hadamard vertex] (15) at (3.75, 1.75) {};
	\node [boundary vertex] (16) at (5.25, 1.75) {};
	\node [green vertex] (17) at (0, 1.5) {};
	\node [green vertex] (18) at (7.5, 1.5) {};
	\node [red vertex] (19) at (9, 1.5) {};
	\node [green vertex] (20) at (10.5, 1.5) {};
	\node [boundary vertex] (21) at (3.25, 1.25) {};
	\node [black vertex] (22) at (-7.5, 1) {};
	\node [black vertex] (23) at (-7, 1) {};
	\node [black vertex] (24) at (-5.5, 1) {};
	\node [black vertex] (25) at (-4, 1) {};
	\node [boundary vertex] (26) at (-0.5, 1) {};
	\node [boundary vertex] (27) at (7, 1) {};
	\node [hadamard vertex] (28) at (8.5, 1) {};
	\node [boundary vertex] (29) at (10, 1) {};
	
	\node [red vertex] (32) at (2.75, 0.75) {};
	
	\node [green vertex] (33) at (5.75, 0.75) {};
	\node [red vertex] (34) at (5.75, 0.75) {};
	\node [boundary vertex] (35) at (8, 0.5) {};
	\node [black vertex] (36) at (-8, 0.25) {};
	\node [black vertex] (37) at (-5, 0.25) {};
	\node [hadamard vertex] (38) at (2.25, 0.25) {};
	\node [hadamard vertex] (39) at (5.25, 0.25) {};
	\node [green vertex] (40) at (-1.5, 0) {};
	\node [red vertex] (41) at (0, 0) {};
	\node [green vertex] (42) at (1.5, 0) {};
	\node [green vertex] (43) at (9, 0) {};
	\node [boundary vertex] (44) at (1.75, -0.25) {};
	\node [boundary vertex] (45) at (4.75, -0.25) {};
	\node [black vertex] (46) at (-9, -0.5) {};
	\node [black vertex] (47) at (-7.5, -0.5) {};
	\node [black vertex] (48) at (-6, -0.5) {};
	\node [black vertex] (49) at (-5.5, -0.5) {};
	\node [boundary vertex] (50) at (-2, -0.5) {};
	\node [hadamard vertex] (51) at (-0.5, -0.5) {};
	\node [boundary vertex] (52) at (1, -0.5) {};
	\node [boundary vertex] (53) at (8.5, -0.5) {};
	
	\node [green vertex] (56) at (2.75, -0.75) {};

	\node [red vertex] (59) at (4.25, -0.75) {};
	\node [green vertex] (60) at (5.75, -0.75) {};
	\node [red vertex] (61) at (5.75, -0.75) {};
	\node [green vertex] (62) at (5.75, -0.75) {};
	\node [boundary vertex] (63) at (-1, -1) {};
	\node [black vertex] (64) at (-8, -1.25) {};
	\node [black vertex] (65) at (-6.5, -1.25) {};
	\node [black vertex] (66) at (-5, -1.25) {};
	\node [boundary vertex] (67) at (2.25, -1.25) {};
	\node [hadamard vertex] (68) at (3.75, -1.25) {};
	\node [boundary vertex] (69) at (5.25, -1.25) {};
	\node [green vertex] (70) at (0, -1.5) {};
	\node [boundary vertex] (71) at (3.25, -1.75) {};
	\node [black vertex] (72) at (-7.5, -2) {};
	\node [boundary vertex] (73) at (-0.5, -2) {};
	\draw  (7) to (15);
	\draw  (53) to (43);
	\draw [out=10, in=190] (70) to (59);
	\draw  (60) to (69);
	\draw  (12) to (1);
	\draw  (46) to (47);
	\draw  (19) to (43);
	\draw  (25) to (24);
	\draw  (9) to (33);
	\draw  (33) to (61);
	\draw  (42) to (33);
	\draw  (7) to (0);
	\draw  (12) to (11);
	\draw  (19) to (20);
	\draw  (24) to (49);
	\draw  (38) to (44);
	\draw  (4) to (14);
	\draw  (36) to (64);
	\draw  (40) to (32);
	\draw  (26) to (17);
	\draw  (59) to (56);
	\draw  (17) to (7);
	\draw  (34) to (39);
	\draw  (41) to (70);
	\draw  (36) to (23);
	\draw  (47) to (22);
	\draw  (23) to (24);
	\draw  (4) to (7);
	\draw  (4) to (32);
	\draw  (9) to (16);
	\draw  (24) to (1);
	\draw  (39) to (45);
	\draw  (11) to (36);
	\draw  (56) to (67);
	\draw  (15) to (21);
	\draw  (65) to (72);
	\draw  (73) to (70);
	\draw  (52) to (42);
	\draw  (61) to (59);
	\draw  (64) to (65);
	\draw  (41) to (42);
	\draw  (47) to (72);
	\draw  (48) to (37);
	\draw  (46) to (36);
	\draw  (37) to (25);
	\draw  (63) to (51);
	\draw  (66) to (37);
	\draw  (33) to (20);
	\draw  (27) to (18);
	\draw  (56) to (32);
	\draw  (22) to (12);
	\draw  (59) to (68);
	\draw  (68) to (71);
	\draw  (37) to (13);
	\draw  (19) to (0);
	\draw  (29) to (20);
	\draw  (32) to (18);
	\draw  (65) to (49);
	\draw  (13) to (12);
	\draw  (48) to (47);
	\draw  (32) to (38);
	\draw  (18) to (19);
	\draw  (28) to (19);
	\draw  (51) to (41);
	\draw  (7) to (9);
	\draw  (2) to (0);
	\draw  (41) to (17);
	\draw  (35) to (28);
	\draw  (50) to (40);
	\draw  (40) to (41);
	\draw  (59) to (43);
	\draw  (65) to (66);
\end{tikzpicture}
\]
\end{minipage}}
   \caption{The 3D cluster state: physical cell geometry and the red/green diagram.} \label{cells}
\end{figure}
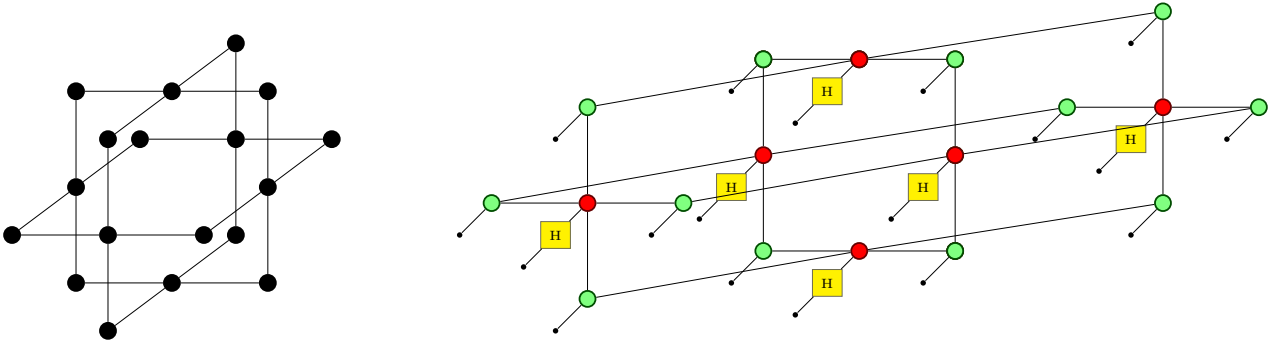
 \begin{figure}[t]
  \centering
    \scalebox{1}{\includegraphics{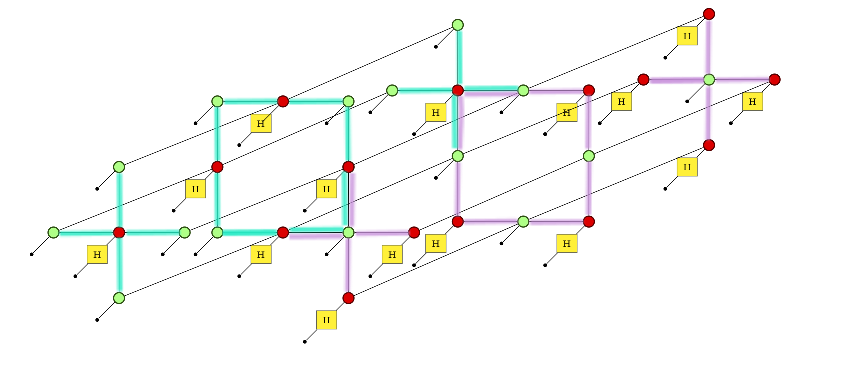}}
   \caption{Primal and dual lattice cells. Primal lattice cells (blue) begin and terminate with red centre-face qubits, and the dual lattice cells (purple) with green qubits.} \label{pdcells}
\end{figure}
 The 3D cluster state in the Raussendorf scheme is made up of basic cells, as shown in figure \ref{cells} along with the associated red/green diagram, with a particular choice of colouring. The `legs' from each qubit represent the process time edges (the reader will appreciate the difficulties of drawing four-dimensional diagrams). The cluster, being tiled from the simple element, is self-similar at a spacing of half a cell; these are known as the \emph{primal} and \emph{dual} lattices. These have a nice representation in the graphical language, as primal lattice cells are the colour-reverse of the dual lattice cells (this emphasises the freedom of choice of a colouring, as long as all choices in the same set of work are compatible). Figure \ref{pdcells} shows both primal and dual cells in the lattice, and how they connect.\\
\indent A particularly nice aspect of representing this cluster in this diagrammatic form is that we can `read off' the lattice stabilizers from the red/green colouring. For example, the set of $X$ operators on each red qubit in figure \ref{cells} is a stabilizer of the cluster. This can be easily verified by inputting a `red $\pi$' node to each existent red node.

\subsection{Defects and measurement}
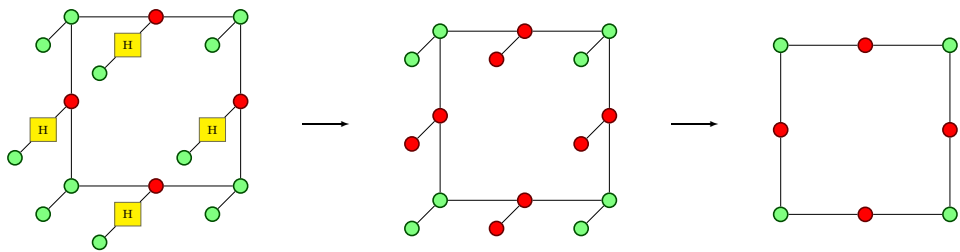
\begin{figure}[t]
\centering
   \scalebox{0.75}{\hspace{-6cm} \begin{minipage}[c]{1.0\linewidth}
  \centering

  \[
\begin{tikzpicture}[quanto]
	\node [green vertex] (0) at (-3, 3) {};
	\node [red vertex] (1) at (0, 3) {};
	\node [green vertex] (2) at (3, 3) {};
	\node [green vertex] (3) at (-4, 2) {};
	\node [hadamard vertex] (4) at (-1, 2) {};
	\node [green vertex] (5) at (2, 2) {};
	\node [green vertex] (6) at (-2, 1) {};
	\node [red vertex] (7) at (-3, 0) {};
	\node [red vertex] (8) at (3, 0) {};
	\node [hadamard vertex] (9) at (-4, -1) {};
	\node [hadamard vertex] (10) at (2, -1) {};
	\node [green vertex] (11) at (-5, -2) {};
	\node [green vertex] (12) at (1, -2) {};
	\node [green vertex] (13) at (-3, -3) {};
	\node [red vertex] (14) at (0, -3) {};
	\node [green vertex] (15) at (3, -3) {};
	\node [green vertex] (16) at (-4, -4) {};
	\node [hadamard vertex] (17) at (-1, -4) {};
	\node [green vertex] (18) at (2, -4) {};
	\node [green vertex] (19) at (-2, -5) {};
	\draw  (8) to (2);
	\draw  (11) to (9);
	\draw  (6) to (4);
	\draw  (1) to (4);
	\draw  (18) to (15);
	\draw  (14) to (17);
	\draw  (8) to (10);
	\draw  (15) to (8);
	\draw  (3) to (0);
	\draw  (7) to (9);
	\draw  (0) to (7);
	\draw  (13) to (14);
	\draw  (16) to (13);
	\draw  (14) to (15);
	\draw  (12) to (10);
	\draw  (0) to (1);
	\draw  (19) to (17);
	\draw  (5) to (2);
	\draw  (1) to (2);
	\draw  (7) to (13);
\end{tikzpicture}
\quad \quad
\rTo
\quad \quad
\begin{tikzpicture}[quanto]
	\node [green vertex] (0) at (-3, 3) {};
	\node [red vertex] (1) at (0, 3) {};
	\node [green vertex] (2) at (3, 3) {};
	\node [green vertex] (3) at (-4, 2) {};
	\node [red vertex] (4) at (-1, 2) {};
	\node [green vertex] (5) at (2, 2) {};
	\node [red vertex] (6) at (-3, 0) {};
	\node [red vertex] (7) at (3, 0) {};
	\node [red vertex] (8) at (-4, -1) {};
	\node [red vertex] (9) at (2, -1) {};
	\node [green vertex] (10) at (-3, -3) {};
	\node [red vertex] (11) at (0, -3) {};
	\node [green vertex] (12) at (3, -3) {};
	\node [green vertex] (13) at (-4, -4) {};
	\node [red vertex] (14) at (-1, -4) {};
	\node [green vertex] (15) at (2, -4) {};
	\draw  (11) to (12);
	\draw  (8) to (6);
	\draw  (3) to (0);
	\draw  (14) to (11);
	\draw  (6) to (10);
	\draw  (5) to (2);
	\draw  (9) to (7);
	\draw  (1) to (2);
	\draw  (4) to (1);
	\draw  (12) to (7);
	\draw  (10) to (11);
	\draw  (13) to (10);
	\draw  (7) to (2);
	\draw  (0) to (1);
	\draw  (15) to (12);
	\draw  (0) to (6);
\end{tikzpicture}
\quad \quad
\rTo
\quad \quad
\begin{tikzpicture}[quanto]
	\node [green vertex] (0) at (-3, 3) {};
	\node [red vertex] (1) at (0, 3) {};
	\node [green vertex] (2) at (3, 3) {};
	\node [red vertex] (3) at (-3, 0) {};
	\node [red vertex] (4) at (3, 0) {};
	\node [green vertex] (5) at (-3, -3) {};
	\node [red vertex] (6) at (0, -3) {};
	\node [green vertex] (7) at (3, -3) {};
	\draw  (4) to (2);
	\draw  (6) to (7);
	\draw  (0) to (3);
	\draw  (1) to (2);
	\draw  (3) to (5);
	\draw  (7) to (4);
	\draw  (5) to (6);
	\draw  (0) to (1);
	\end{tikzpicture}
\]
\end{minipage}}
\caption{The result of measurement in the $X$ basis in the cluster. The process graph now becomes topologically identical with the physical cluster.}
\label{Xmment}
\end{figure}

Logical qubits are defined in the cluster state by defects, measuring out the central qubit in the face of a line of successive cluster cells. Figure \ref{defect} shows the effect on one cell of producing a defect; as we know, the measurement has the effect of removing the measured qubit from the lattice, and this is shown in the corresponding diagram. Figure \ref{defect} shows the measurement on one face (the final step in an application of the re-write rule of figure \ref{combin}(c)), and figure \ref{clusterdefect} shows the whole cell.\\
 \begin{figure}[t]
   \centering
    \scalebox{1}{ \begin{minipage}[c]{1.0\linewidth}
  \centering

  \[
\begin{tikzpicture}[quanto]
\node [red vertex] (centre) at (1.5,1.5) {};
\node [green vertex] (a) at (1.5,0.0) {};
\node [green vertex] (b) at (1.5,3.0) {};
\node [green vertex] (c) at (0.0,1.5) {};
\node [green vertex] (d) at (3.0,1.5) {};
\draw [] (a) to (centre);
\draw [] (b) to (centre);
\draw [] (c) to (centre);
\draw [] (d) to (centre);
\node [boundary vertex] (ba) at (1.0,-0.5) {};
\draw [] (a) to (ba);
\node [boundary vertex] (bb) at (1.0,2.5) {};
\draw [] (b) to (bb);
\node [boundary vertex] (bc) at (-0.5,1.0) {};
\draw [] (c) to (bc);
\node [boundary vertex] (bd) at (2.5,1.0) {};
\draw [] (d) to (bd);
\node [hadamard vertex] (h) at (0.75,0.75) {};
\node [red vertex] (bh) at (0,0) {};
\draw [] (centre) to (h);
\draw [] (h) to (bh);
\end{tikzpicture}
\rTo
\begin{tikzpicture}[quanto]
\node [red vertex] (centre) at (1.5,1.5) {};
\node [green vertex] (a) at (1.5,0.0) {};
\node [green vertex] (b) at (1.5,3.0) {};
\node [green vertex] (c) at (0.0,1.5) {};
\node [green vertex] (d) at (3.0,1.5) {};
\draw [] (a) to (centre);
\draw [] (b) to (centre);
\draw [] (c) to (centre);
\draw [] (d) to (centre);
\node [boundary vertex] (ba) at (1.0,-0.5) {};
\draw [] (a) to (ba);
\node [boundary vertex] (bb) at (1.0,2.5) {};
\draw [] (b) to (bb);
\node [boundary vertex] (bc) at (-0.5,1.0) {};
\draw [] (c) to (bc);
\node [boundary vertex] (bd) at (2.5,1.0) {};
\draw [] (d) to (bd);
\node [green vertex] (h) at (0.75,0.75) {};
\draw [] (centre) to (h);
\end{tikzpicture}
\rTo
\begin{tikzpicture}[quanto]
\node [green vertex] (a) at (1.5,0.0) {};
\node [green vertex] (b) at (1.5,3.0) {};
\node [green vertex] (c) at (0.0,1.5) {};
\node [green vertex] (d) at (3.0,1.5) {};
\node [boundary vertex] (ba) at (1.0,-0.5) {};
\draw [] (a) to (ba);
\node [boundary vertex] (bb) at (1.0,2.5) {};
\draw [] (b) to (bb);
\node [boundary vertex] (bc) at (-0.5,1.0) {};
\draw [] (c) to (bc);
\node [boundary vertex] (bd) at (2.5,1.0) {};
\draw [] (d) to (bd);
\end{tikzpicture}
\]
\end{minipage}} 
      \caption{Measuring out one face qubit in a cluster cell.} \label{defect}
\end{figure}
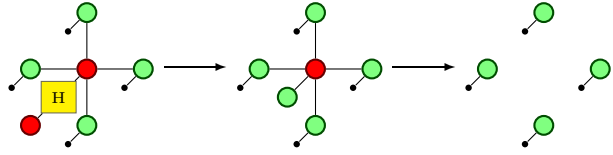
 \begin{figure}[t]
   \centering
    \scalebox{0.6}{\hspace{-14cm} \begin{minipage}[c]{1.0\linewidth}
  \centering

  \[
\begin{tikzpicture}
	\node [green vertex] (0) at (4.25, 5) {};
		\node [angle] at (0) {Z};
	\node [boundary vertex] (1) at (3.75, 4.5) {};
	\node [green vertex] (2) at (-2.5, 3.5) {};
	\node [red vertex] (3) at (-0.5, 3.5) {};
		\node [angle] at (3) {X};
	\node [green vertex] (4) at (1.5, 3.5) {};
	\node [boundary vertex] (5) at (-3, 3) {};
	\node [hadamard vertex] (6) at (-1, 3) {};
	\node [boundary vertex] (7) at (0.75, 3) {};
	\node [green vertex] (8) at (2.25, 3) {};
		\node [angle] at (8) {Z};
	\node [green vertex] (9) at (6.25, 3) {};
		\node [angle] at (9) {Z};
	\node [boundary vertex] (10) at (-1.5, 2.5) {};
	\node [boundary vertex] (11) at (1.75, 2.5) {};
	\node [boundary vertex] (12) at (5.75, 2.5) {};
	\node [green vertex] (13) at (-5, 2) {};
	\node [boundary vertex] (14) at (-5.5, 1.5) {};
	\node [red vertex] (15) at (-2.5, 1.5) {};
		\node [angle] at (15) {X};
	\node [red vertex] (16) at (1.5, 1.5) {};
		\node [angle] at (16) {X};
	\node [hadamard vertex] (17) at (-3, 1) {};
	\node [hadamard vertex] (18) at (1, 1) {};
	\node [green vertex] (19) at (4.25, 1) {};
		\node [angle] at (19) {Z};
	\node [boundary vertex] (20) at (-3.5, 0.5) {};
	\node [boundary vertex] (21) at (0.5, 0.5) {};
	\node [boundary vertex] (22) at (3.75, 0.5) {};
	\node [green vertex] (23) at (-11.75, 0.25) {};
	\node [red vertex] (24) at (-9.75, 0.25) {};
		\node [angle] at (24) {X};
	\node [green vertex] (25) at (-7.75, 0.25) {};
	\node [green vertex] (26) at (-7, 0) {};
	\node [green vertex] (27) at (-3, 0) {};
	\node [boundary vertex] (28) at (-12.25, -0.25) {};
	\node [hadamard vertex] (29) at (-10.25, -0.25) {};
	\node [boundary vertex] (30) at (-8.5, -0.25) {};
	\node [boundary vertex] (31) at (-7.5, -0.5) {};
	\node [boundary vertex] (32) at (-3.5, -0.5) {};
	\node [green vertex] (33) at (-2.5, -0.5) {};
	\node [red vertex] (34) at (-0.5, -0.5) {};
		\node [angle] at (34) {X};
	\node [green vertex] (35) at (1.5, -0.5) {};
	\node [boundary vertex] (36) at (-10.75, -0.75) {};
	\node [boundary vertex] (37) at (-3, -1) {};
	\node [hadamard vertex] (38) at (-1, -1) {};
	\node [boundary vertex] (39) at (1, -1) {};
	\node [green vertex] (40) at (-15.5, -1.5) {};%
		\node [angle] at (40) {Z};
	\node [boundary vertex] (41) at (-1.5, -1.5) {};
	\node [red vertex] (42) at (-11.75, -1.75) {};
		\node [angle] at (42) {X};
	\node [red vertex] (43) at (-7.75, -1.75) {};
		\node [angle] at (43) {X};
	\node [boundary vertex] (44) at (-16, -2) {};
	\node [green vertex] (45) at (-5, -2) {};
	\node [hadamard vertex] (46) at (-12.25, -2.25) {};
	\node [hadamard vertex] (47) at (-8.25, -2.25) {};
	\node [boundary vertex] (48) at (-5.5, -2.5) {};
	\node [boundary vertex] (49) at (-12.75, -2.75) {};
	\node [boundary vertex] (50) at (-8.75, -2.75) {};
	\node [green vertex] (51) at (-17.5, -3.5) {};%
		\node [angle] at (51) {Z};
	\node [green vertex] (52) at (-13.5, -3.5) {};%
		\node [angle] at (52) {Z};
	\node [green vertex] (53) at (-11.75, -3.75) {};
	\node [red vertex] (54) at (-9.75, -3.75) {};
		\node [angle] at (54) {X};
	\node [green vertex] (55) at (-7.75, -3.75) {};
	\node [boundary vertex] (56) at (-18, -4) {};
	\node [boundary vertex] (57) at (-14, -4) {};
	\node [boundary vertex] (58) at (-12.25, -4.25) {};
	\node [hadamard vertex] (59) at (-10.25, -4.25) {};
	\node [boundary vertex] (60) at (-8.25, -4.25) {};
	\node [boundary vertex] (61) at (-10.75, -4.75) {};
	\node [green vertex] (62) at (-15.5, -5.5) {};%
		\node [angle] at (62) {Z};
	\node [boundary vertex] (63) at (-16, -6) {};
	\draw  (55) to (60);
	\draw  (13) to (14);
	\draw  (45) to (48);
	\draw  (19) to (22);
	\draw  (3) to (0);
	\draw  (62) to (54);
	\draw  (43) to (25);
	\draw  (27) to (32);
	\draw  (9) to (12);
	\draw  (3) to (4);
	\draw  (13) to (3);
	\draw  (54) to (59);
	\draw  (16) to (18);
	\draw  (15) to (33);
	\draw  (26) to (15);
	\draw  (29) to (36);
	\draw  (43) to (27);
	\draw  (0) to (1);
	\draw  (47) to (50);
	\draw  (46) to (49);
	\draw  (26) to (31);
	\draw  (2) to (5);
	\draw  (34) to (38);
	\draw  (23) to (28);
	\draw  (15) to (8);
	\draw  (34) to (19);
	\draw  (27) to (16);
	\draw  (33) to (34);
	\draw  (62) to (63);
	\draw  (23) to (24);
	\draw  (59) to (61);
	\draw  (42) to (26);
	\draw  (40) to (44);
	\draw  (43) to (47);
	\draw  (24) to (29);
	\draw  (24) to (25);
	\draw  (24) to (13);
	\draw  (4) to (7);
	\draw  (6) to (10);
	\draw  (45) to (34);
	\draw  (16) to (4);
	\draw  (2) to (15);
	\draw  (38) to (41);
	\draw  (52) to (43);
	\draw  (16) to (9);
	\draw  (54) to (45);
	\draw  (35) to (39);
	\draw  (34) to (35);
	\draw  (18) to (21);
	\draw  (40) to (24);
	\draw  (35) to (16);
	\draw  (42) to (46);
	\draw  (23) to (42);
	\draw  (42) to (53);
	\draw  (53) to (58);
	\draw  (33) to (37);
	\draw  (55) to (43);
	\draw  (51) to (42);
	\draw  (15) to (17);
	\draw  (2) to (3);
	\draw  (3) to (6);
	\draw  (17) to (20);
	\draw  (8) to (11);
	\draw  (54) to (55);
	\draw  (53) to (54);
	\draw  (25) to (30);
	\draw  (51) to (56);
	\draw  (52) to (57);
	\end{tikzpicture}
\]
\end{minipage}} 
      \caption{Creating a defect through the cluster.} \label{clusterdefect}
\end{figure}
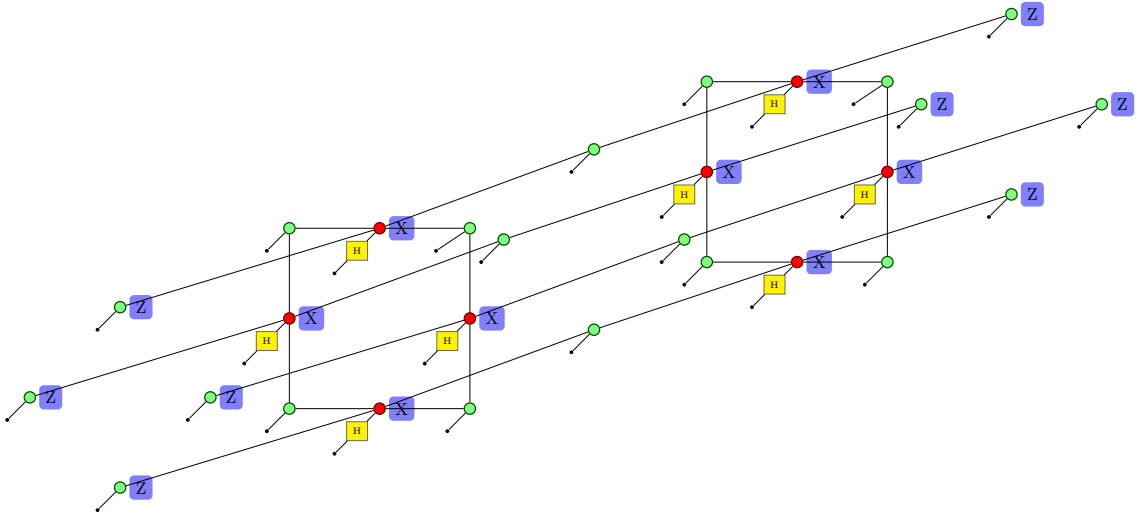
\indent We can see from this diagram that we are left with a ``tube" of four qubits at each crossection of the lattice, alternating red and green. Again it is instructive to consider the stabilizers for this tube: $Z$ on the green face nodes at the beginning and end of the tube, and $X$ on red nodes on the intervening cells, as marked on figure \ref{clusterdefect}. This is indeed exactly the \emph{correlation surface} surrounding a defect, and the entity that in fact encodes a logical qubit in the model. Each qubit in the surface is connected to one at an ``earlier computational time", and one at a ``later computational time". The four qubits in the centre of each cell are connected together by four other qubits into a ring. If we create a larger defect by measuring out more qubits, this ring becomes larger and connects together more than four `lines' of physical qubits into the correlation surface.\\
\indent With the physical qubits within the defects measured out in the $Z$ basis, computation proceeds by measuring all other qubits in the $X$ basis. The outcome of this on one part of a cell (post-selected on the result of the stabilizer measurement; see \S\ref{error-correction}) is shown in figure \ref{Xmment}; it is trivial to generalise this. This is an extremely important result: \emph{the action of the measurement reduces the 4D open process graph to a 3D closed process graph that is topologically isomorphic to the physical cluster and defect}. 

We will see that this is the key factor in using the pictorial language for describing logical defects. 

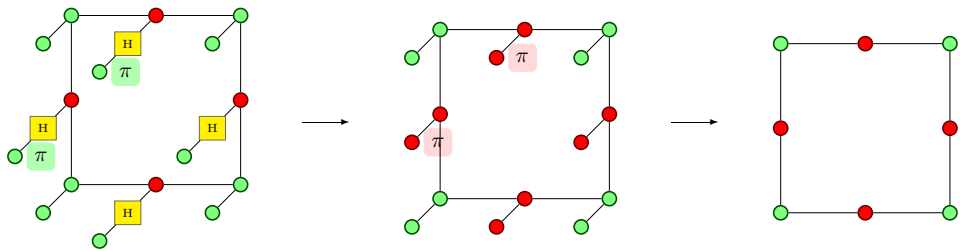
\begin{figure}[t]
\centering
   \scalebox{0.75}{\hspace{-6cm} \begin{minipage}[c]{1.0\linewidth}
  \centering

  \[
\begin{tikzpicture}[quanto]
	\node [green vertex] (0) at (-3, 3) {};
	\node [red vertex] (1) at (0, 3) {};
	\node [green vertex] (2) at (3, 3) {};
	\node [green vertex] (3) at (-4, 2) {};
	\node [hadamard vertex] (4) at (-1, 2) {};
	\node [green vertex] (5) at (2, 2) {};
	\node [green vertex] (6) at (-2, 1) {};
	\node [green angle] at (6) {$\pi$};
	
	\node [red vertex] (7) at (-3, 0) {};
	\node [red vertex] (8) at (3, 0) {};
	\node [hadamard vertex] (9) at (-4, -1) {};
	\node [hadamard vertex] (10) at (2, -1) {};
	\node [green vertex] (11) at (-5, -2) {};
	\node [green angle] at (11) {$\pi$};
	
	\node [green vertex] (12) at (1, -2) {};
	\node [green vertex] (13) at (-3, -3) {};
	\node [red vertex] (14) at (0, -3) {};
	\node [green vertex] (15) at (3, -3) {};
	\node [green vertex] (16) at (-4, -4) {};
	\node [hadamard vertex] (17) at (-1, -4) {};
	\node [green vertex] (18) at (2, -4) {};
	\node [green vertex] (19) at (-2, -5) {};
	\draw  (8) to (2);
	\draw  (11) to (9);
	\draw  (6) to (4);
	\draw  (1) to (4);
	\draw  (18) to (15);
	\draw  (14) to (17);
	\draw  (8) to (10);
	\draw  (15) to (8);
	\draw  (3) to (0);
	\draw  (7) to (9);
	\draw  (0) to (7);
	\draw  (13) to (14);
	\draw  (16) to (13);
	\draw  (14) to (15);
	\draw  (12) to (10);
	\draw  (0) to (1);
	\draw  (19) to (17);
	\draw  (5) to (2);
	\draw  (1) to (2);
	\draw  (7) to (13);
\end{tikzpicture}
\quad \quad
\rTo
\quad \quad
\begin{tikzpicture}[quanto]
	\node [green vertex] (0) at (-3, 3) {};
	\node [red vertex] (1) at (0, 3) {};
	\node [green vertex] (2) at (3, 3) {};
	\node [green vertex] (3) at (-4, 2) {};
	\node [red vertex] (4) at (-1, 2) {};
	\node [red angle] at (4) {$\pi$};
	
	\node [green vertex] (5) at (2, 2) {};
	\node [red vertex] (6) at (-3, 0) {};
	\node [red vertex] (7) at (3, 0) {};
	\node [red vertex] (8) at (-4, -1) {};
	\node [red angle] at (8) {$\pi$};
	
	\node [red vertex] (9) at (2, -1) {};
	\node [green vertex] (10) at (-3, -3) {};
	\node [red vertex] (11) at (0, -3) {};
	\node [green vertex] (12) at (3, -3) {};
	\node [green vertex] (13) at (-4, -4) {};
	\node [red vertex] (14) at (-1, -4) {};
	\node [green vertex] (15) at (2, -4) {};
	\draw  (11) to (12);
	\draw  (8) to (6);
	\draw  (3) to (0);
	\draw  (14) to (11);
	\draw  (6) to (10);
	\draw  (5) to (2);
	\draw  (9) to (7);
	\draw  (1) to (2);
	\draw  (4) to (1);
	\draw  (12) to (7);
	\draw  (10) to (11);
	\draw  (13) to (10);
	\draw  (7) to (2);
	\draw  (0) to (1);
	\draw  (15) to (12);
	\draw  (0) to (6);
\end{tikzpicture}
\quad \quad
\rTo
\quad \quad
\begin{tikzpicture}[quanto]
	\node [green vertex] (0) at (-3, 3) {};
	\node [red vertex] (1) at (0, 3) {};
	\node [green vertex] (2) at (3, 3) {};
	\node [red vertex] (3) at (-3, 0) {};
	\node [red vertex] (4) at (3, 0) {};
	\node [green vertex] (5) at (-3, -3) {};
	\node [red vertex] (6) at (0, -3) {};
	\node [green vertex] (7) at (3, -3) {};
	\draw  (4) to (2);
	\draw  (6) to (7);
	\draw  (0) to (3);
	\draw  (1) to (2);
	\draw  (3) to (5);
	\draw  (7) to (4);
	\draw  (5) to (6);
	\draw  (0) to (1);
	\end{tikzpicture}
\]
\end{minipage}}
\caption{Measurement of the cluster stabilizer. An even number (here two) of the individual red qubits have the measurement outcome -1 (``green $\pi$"), but the diagram re-writes to the same normal form as figure \ref{Xmment}. }
\label{Pmment}
\end{figure}

\subsection{Error correction}\label{error-correction}

We have noted that, in order to perform the re-writes in figure \ref{Xmment}, we post-select on measurement results. What is meant by this is that the re-write as given is formally a short-hand for a more complicated graphical calculation involving the interplay of quantum and classical information. For full details, the reader is directed to \cite[\S3.3.4 and \S12]{monster}. The convention is to represent the measurement in the diagrams as giving the `+1' outcome in the relevant basis; we now need to show that the diagrams we have thus derived are fully general for all potential measurement outcomes.

In the present scheme, as is the case in standard measurement-based computation, the state of the system may be changed given the outcomes of measurements. In topological cluster-state computing, the $X$-basis measurements that occur at each step of computational time should, in the case of no cluster  errors, give a sequence of results of a given parity  \cite[\S VII]{austin3D}. The six qubits in the centre of each face of a cell should measure out to give the combined parity of the measurements as +1. If we consider the primal (blue) cell in figure  \ref{pdcells} then that parity requirement means that an even number of the six red face qubits are measured with ``green  $\pi$" (-1 outcome) rather than simply a green node (+1 outcome) in figure  \ref{Xmment}. It is easy to see that the subsequent even number of ``red  $\pi$" nodes in the final diagram will cancel, leaving the same post-measurement diagram, figure  \ref{Pmment}.

There is therefore a single normal form diagram for the ideal cluster after measurement,  regardless of the results of the individual qubit measurements. The error correction scheme of the topological model uses the results of the measurement to find when cluster qubits deviate from the correct state. Corrections are then applied that return the state to the desired eigenstate (see for example \cite[\S VII]{austin3D}). We need therefore to remember when using the red/green calculus that re-writes of the form figure \ref{Xmment} occur based on the used of active error correction techniques.

\subsection{Creating a logical qubit calculus}
\begin{figure}[t]
\centering
 \hspace{-4cm}   \scalebox{0.75}{\hspace{-6cm} \begin{minipage}[c]{1.0\linewidth}
  \centering

  \[
\begin{tikzpicture}[quanto]
	\node [green vertex] (0) at (-2, 1.5) {};
		\node [boundary vertex] (a) at (-2,2.5) {};
		\draw [] (a) to (0);
	\node [green vertex] (1) at (0, 1.5) {};
		\node [boundary vertex] (b) at (0,2.5) {};
		\draw [] (b) to (1);
	\node [green vertex] (2) at (2, 1.5) {};
		\node [boundary vertex] (c) at (2,2.5) {};
		\draw [] (c) to (2);
	\node [green vertex] (3) at (4, 1.5) {};
		\node [boundary vertex] (d) at (4,2.5) {};
		\draw [] (d) to (3);
	\node [green vertex] (4) at (-3, 0.5) {};
	\node [red vertex] (5) at (4, 0.5) {};
	\node [red vertex] (6) at (-2, 0) {};
	\node [green vertex] (7) at (-1, 0) {};
	\node [red vertex] (8) at (0, 0) {};
	\node [green vertex] (9) at (1, 0) {};
	\node [red vertex] (10) at (2, 0) {};
	\node [green vertex] (11) at (3, 0) {};
	\node [green vertex] (12) at (-2, -1) {};
		\node [boundary vertex] (e) at (-2,-2) {};
		\draw [] (e) to (12);
	\node [green vertex] (13) at (0, -1) {};
		\node [boundary vertex] (f) at (0,-2) {};
		\draw [] (f) to (13);
	\node [green vertex] (14) at (2, -1) {};
		\node [boundary vertex] (g) at (2,-2) {};
		\draw [] (g) to (14);
	\node [green vertex] (15) at (4, -1) {};
		\node [boundary vertex] (h) at (4,-2) {};
		\draw [] (h) to (15);
	\draw  (1) to (8);
	\draw  (6) to (12);
	\draw  (0) to (6);
	\draw  (2) to (10);
	\draw  (11) to (5);
	\draw  (6) to (7);
	\draw  (9) to (10);
	\draw  (10) to (14);
	\draw  (4) to (6);
	\draw  (5) to (4);
	\draw  (3) to (5);
	\draw  (8) to (9);
	\draw  (8) to (13);
	\draw  (5) to (15);
	\draw  (10) to (11);
	\draw  (7) to (8);
\end{tikzpicture}
\quad \quad
\rTo
\quad \quad
\begin{tikzpicture}[quanto]
	\node [green vertex] (0) at (-2, 1.5) {};
		\node [boundary vertex] (a) at (-2,2.5) {};
		\draw [] (a) to (0);
	\node [green vertex] (1) at (0, 1.5) {};
		\node [boundary vertex] (b) at (0,2.5) {};
		\draw [] (b) to (1);
	\node [green vertex] (2) at (2, 1.5) {};
		\node [boundary vertex] (c) at (2,2.5) {};
		\draw [] (c) to (2);
	\node [green vertex] (3) at (4, 1.5) {};
		\node [boundary vertex] (d) at (4,2.5) {};
		\draw [] (d) to (3);
	\node [red vertex] (4) at (-2, 0.5) {};
	\node [red vertex] (5) at (4, 0.5) {};
	\node [red vertex] (6) at (0, 0) {};
	\node [red vertex] (7) at (2, 0) {};
	\node [green vertex] (8) at (-2, -1) {};
		\node [boundary vertex] (e) at (-2,-2) {};
		\draw [] (e) to (8);
	\node [green vertex] (9) at (0, -1) {};
		\node [boundary vertex] (f) at (0,-2) {};
		\draw [] (f) to (9);
	\node [green vertex] (10) at (2, -1) {};
		\node [boundary vertex] (g) at (2,-2) {};
		\draw [] (g) to (10);
	\node [green vertex] (11) at (4, -1) {};
		\node [boundary vertex] (h) at (4,-2) {};
		\draw [] (h) to (11);
	\draw  (1) to (6);
	\draw  (4) to (8);
	\draw  (0) to (4);
	\draw  (2) to (7);
	\draw  (7) to (10);
	\draw  (4) to (5);
	\draw  (5) to (7);
	\draw  (6) to (4);
	\draw  (3) to (5);
	\draw  (7) to (6);
	\draw  (6) to (9);
	\draw  (5) to (11);
\end{tikzpicture}
\quad \quad
\rTo
\quad \quad
\begin{tikzpicture}[quanto]
	\node [green vertex] (0) at (-2, 1) {};
		\node [boundary vertex] (a) at (-2,2) {};
		\draw [] (a) to (0);
	\node [green vertex] (1) at (0, 1) {};
		\node [boundary vertex] (b) at (0,2) {};
		\draw [] (b) to (1);
	\node [green vertex] (2) at (2, 1) {};
		\node [boundary vertex] (c) at (2,2) {};
		\draw [] (c) to (2);
	\node [green vertex] (3) at (4, 1) {};
		\node [boundary vertex] (d) at (4,2) {};
		\draw [] (d) to (3);
	\node [red vertex] (4) at (1, 0) {};
	\node [green vertex] (5) at (-2, -1) {};
		\node [boundary vertex] (a) at (-2,-2) {};
		\draw [] (e) to (5);
	\node [green vertex] (6) at (0, -1) {};
		\node [boundary vertex] (f) at (0,-2) {};
		\draw [] (f) to (6);
	\node [green vertex] (7) at (2, -1) {};
		\node [boundary vertex] (g) at (2,-2) {};
		\draw [] (g) to (7);
	\node [green vertex] (8) at (4, -1) {};
		\node [boundary vertex] (h) at (4,-2) {};
		\draw [] (h) to (8);
	\draw  (4) to (5);
	\draw  (4) to (7);
	\draw  (4) to (8);
	\draw  (0) to (4);
	\draw  (2) to (4);
	\draw  (3) to (4);
	\draw  (4) to (1);
	\draw  (4) to (6);
	\end{tikzpicture}
	\quad \quad
\rTo
\quad \quad
\begin{tikzpicture}[quanto]
	\node [green vertex] (0) at (-1, 2) {};
		\node [boundary vertex] (a) at (-1,3) {};
		\draw [] (a) to (0);
	\node [green vertex] (1) at (0, 2) {};
		\node [boundary vertex] (b) at (0,3) {};
		\draw [] (b) to (1);
	\node [green vertex] (2) at (1, 2) {};
		\node [boundary vertex] (c) at (1,3) {};
		\draw [] (c) to (2);
	\node [green vertex] (3) at (2, 2) {};
		\node [boundary vertex] (d) at (2,3) {};
		\draw [] (d) to (3);
	\node [red vertex] (4) at (0.5, 1) {};
	\node [red vertex] (5) at (0.5, 0) {};
	\node [green vertex] (6) at (-1, -1) {};
		\node [boundary vertex] (e) at (-1,-2) {};
		\draw [] (e) to (6);
	\node [green vertex] (7) at (0, -1) {};
		\node [boundary vertex] (f) at (0,-2) {};
		\draw [] (f) to (7);
	\node [green vertex] (8) at (1, -1) {};
		\node [boundary vertex] (g) at (1,-2) {};
		\draw [] (g) to (8);
	\node [green vertex] (9) at (2, -1) {};
		\node [boundary vertex] (h) at (2,-2) {};
		\draw [] (h) to (9);
	\draw  (5) to (4);
	\draw  (6) to (5);
	\draw  (0) to (4);
	\draw  (2) to (4);
	\draw  (8) to (5);
	\draw  (9) to (5);
	\draw  (3) to (4);
	\draw  (7) to (5);
	\draw  (4) to (1);
\end{tikzpicture}
\]
\end{minipage}}
\caption{Using the copying and then spider rules to create a single logical qubit line from a cluster defect.}
\label{spider}
\end{figure}
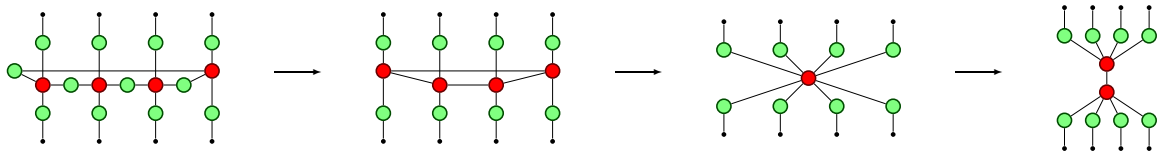

We can now use this result to look at how the topology of the defect strands is represented within the graphical language. In order to show that the geometries of defect strands and red/green graphs are identical, we will construct the lattice with defects in the calculus, and then demonstrate the structure re-writes to a single logical qubit line per defect that follows the geometry of that defect. We will also show that process time and computational time now agree. We will show how primal and dual lattices behave, and construct single-qubit operators. Finally we will construct a CNOT gate from the calculus and show that it agrees with the CNOT used in the topological model.

We start by applying the red/green rewrite rules to the cluster diagram with defect, figure \ref{clusterdefect}. Figure \ref{spider} shows the key result from re-writing a single cell. The graph layout has been changed to make the rewrites easier to discern: the four green vertices at the top and bottom are the four face vertices of a cell. Note that we have only been able to apply the re-write rules in this form on the three-dimensional graph rather than the initial four-dimensional one, showing how key the result of the previous section is. The final result from the rewrites is a single logical qubit line, along the same geometry as the physical defect. The top-bottom process time line is now isomorphic with the computational time into the depth of the lattice. The most important rewrite tool used is the spider rule -- it then becomes simple to see that any number of qubit lines will collapse into a single logical line centred on the defect. A simple extension of this, using the spider rule again, makes it evident that two defects in an entangled pair, as is used to represent one logical qubit in topological cluster-state QC, will also collapse to a single qubit line.

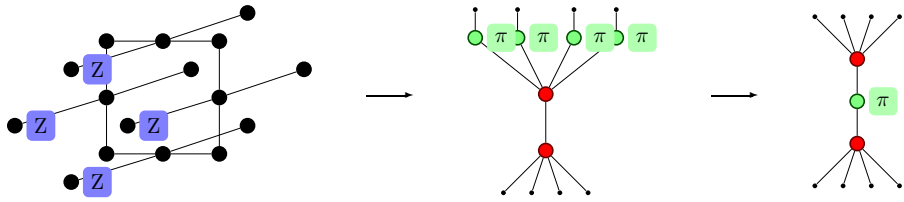
\begin{figure}[t]
\centering
     \scalebox{0.75}{\hspace{-6cm} \begin{minipage}[c]{1.0\linewidth}
  \centering

  \[
\begin{tikzpicture}[quanto]
	\node [black vertex] (0) at (3, 3) {};
	\node [black vertex] (1) at (-2, 2) {};
	\node [black vertex] (2) at (0, 2) {};
	\node [black vertex] (3) at (2, 2) {};
	\node [black vertex] (4) at (-3.25, 1) {};
	\node [black vertex] (5) at (1, 1) {};
	\node [black vertex] (6) at (5, 1) {};
	\node [black vertex] (7) at (-2, 0) {};
	\node [black vertex] (8) at (2, 0) {};
	\node [black vertex] (9) at (-5.25, -1) {};
	\node [black vertex] (10) at (-1.25, -1) {};
	\node [black vertex] (11) at (3, -1) {};
	\node [black vertex] (12) at (-2, -2) {};
	\node [black vertex] (13) at (0, -2) {};
	\node [black vertex] (14) at (2, -2) {};
	\node [black vertex] (15) at (-3.25, -3) {};
	\draw  (13) to (14);
	\draw  (3) to (2);
	\draw  (13) to (15);
	\draw  (7) to (12);
	\draw  (13) to (11);
	\draw  (12) to (13);
	\draw  (1) to (7);
	\draw  (8) to (6);
	\draw  (8) to (3);
	\draw  (14) to (8);
	\draw  (2) to (0);
	\draw  (10) to (8);
	\draw  (7) to (5);
	\draw  (2) to (1);
	\draw  (9) to (7);
	\draw  (4) to (2);
	\node [angle] at (15) {Z};
	\node [angle] at (10) {Z};
	\node [angle] at (9) {Z};
	\node [angle] at (4) {Z};
\end{tikzpicture}
\quad \quad
\rTo
\quad \quad
\begin{tikzpicture}[quanto]
	\node [boundary vertex] (0) at (-2.5, 4) {};
	\node [boundary vertex] (1) at (-1, 4) {};
	\node [boundary vertex] (2) at (1, 4) {};
	\node [boundary vertex] (3) at (2.5, 4) {};
	\node [green vertex] (4) at (-2.5, 3) {};
	\node [green vertex] (5) at (-1, 3) {};
	\node [green vertex] (6) at (1, 3) {};
	\node [green vertex] (7) at (2.5, 3) {};
	\node [red vertex] (8) at (0, 1) {};
	\node [red vertex] (9) at (0, -1) {};
	\node [boundary vertex] (10) at (-1.5, -2.5) {};
	\node [boundary vertex] (11) at (-0.5, -2.5) {};
	\node [boundary vertex] (12) at (0.5, -2.5) {};
	\node [boundary vertex] (13) at (1.5, -2.5) {};
	\draw  (9) to (10);
	\draw  (9) to (11);
	\draw  (9) to (13);
	\draw  (5) to (8);
	\draw  (12) to (9);
	\draw  (9) to (8);
	\draw  (1) to (5);
	\draw  (7) to (3);
	\draw  (6) to (8);
	\draw  (7) to (8);
	\draw  (4) to (8);
	\draw  (4) to (0);
	\draw  (6) to (2);
	\node [green angle] at (4) {$\pi$};
	\node [green angle] at (5) {$\pi$};
	\node [green angle] at (6) {$\pi$};
	\node [green angle] at (7) {$\pi$};
\end{tikzpicture}
\quad \quad
\rTo
\quad \quad
\begin{tikzpicture}[quanto]
	\node [boundary vertex] (0) at (-1.5, 3.5) {};
	\node [boundary vertex] (1) at (-0.5, 3.5) {};
	\node [boundary vertex] (2) at (0.5, 3.5) {};
	\node [boundary vertex] (3) at (1.5, 3.5) {};
	\node [red vertex] (4) at (0, 2) {};
	\node [green vertex] (5) at (0, 0.5) {};
	\node [red vertex] (6) at (0, -1) {};
	\node [boundary vertex] (7) at (-1.5, -2.5) {};
	\node [boundary vertex] (8) at (-0.5, -2.5) {};
	\node [boundary vertex] (9) at (0.5, -2.5) {};
	\node [boundary vertex] (10) at (1.5, -2.5) {};
	\draw  (4) to (1);
	\draw  (5) to (4);
	\draw  (6) to (10);
	\draw  (2) to (4);
	\draw  (6) to (7);
	\draw  (9) to (6);
	\draw  (4) to (0);
	\draw  (6) to (5);
	\draw  (4) to (3);
	\draw  (6) to (8);
	\node [green angle] at (5) {$\pi$};
\end{tikzpicture}
\]
\end{minipage}}
\caption{A ring of physical $Z$ operations becomes the logical $Z_L$ operator.}
\label{ring}
\end{figure}

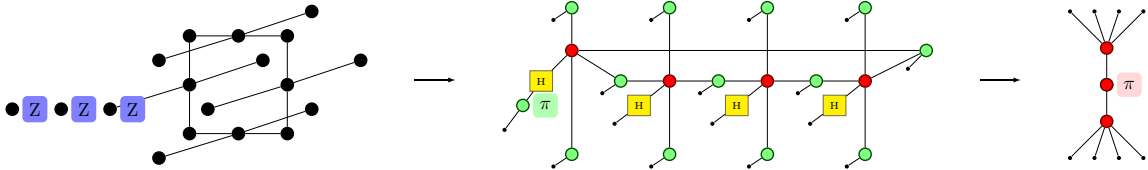
\begin{figure}[t]
\centering
  \hspace{-4cm}   \scalebox{0.65}{\hspace{-10cm} \begin{minipage}[c]{1.0\linewidth}
  \centering

  \[
\begin{tikzpicture}[quanto]
	\node [black vertex] (0) at (3, 3) {};
	\node [black vertex] (1) at (-2, 2) {};
	\node [black vertex] (2) at (0, 2) {};
	\node [black vertex] (3) at (2, 2) {};
	\node [black vertex] (4) at (-3.25, 1) {};
	\node [black vertex] (5) at (1, 1) {};
	\node [black vertex] (6) at (5, 1) {};
	\node [black vertex] (7) at (-2, 0) {};
	\node [black vertex] (8) at (2, 0) {};
	\node [black vertex] (9) at (-5.25, -1) {};
		\node [black vertex] (16) at (-7.25, -1) {};
		\node [black vertex] (17) at (-9.25, -1) {};
	\node [black vertex] (10) at (-1.25, -1) {};
	\node [black vertex] (11) at (3, -1) {};
	\node [black vertex] (12) at (-2, -2) {};
	\node [black vertex] (13) at (0, -2) {};
	\node [black vertex] (14) at (2, -2) {};
	\node [black vertex] (15) at (-3.25, -3) {};
	\draw  (13) to (14);
	\draw  (3) to (2);
	\draw  (13) to (15);
	\draw  (7) to (12);
	\draw  (13) to (11);
	\draw  (12) to (13);
	\draw  (1) to (7);
	\draw  (8) to (6);
	\draw  (8) to (3);
	\draw  (14) to (8);
	\draw  (2) to (0);
	\draw  (10) to (8);
	\draw  (7) to (5);
	\draw  (2) to (1);
	\draw  (9) to (7);
	\draw  (4) to (2);
	\node [angle] at (9) {Z};
	\node [angle] at (16) {Z};
	\node [angle] at (17) {Z};
\end{tikzpicture}
\quad \quad
\rTo
\quad \quad
\begin{tikzpicture}[quanto]
	\node [green vertex] (0) at (-4, 3) {};
	\node [green vertex] (1) at (0, 3) {};
	\node [green vertex] (2) at (4, 3) {};
	\node [green vertex] (3) at (8, 3) {};
	\node [boundary vertex] (4) at (-4.75, 2.5) {};
	\node [boundary vertex] (5) at (-0.75, 2.5) {};
	\node [boundary vertex] (6) at (3.25, 2.5) {};
	\node [boundary vertex] (7) at (7.25, 2.5) {};
	\node [red vertex] (8) at (-4, 1.25) {};
	\node [green vertex] (9) at (10.5, 1.25) {};
	\node [boundary vertex] (10) at (9.75, 0.5) {};
	\node [hadamard vertex] (11) at (-5.25, 0) {};
	\node [green vertex] (12) at (-2, 0) {};
	\node [red vertex] (13) at (0, 0) {};
	\node [green vertex] (14) at (2, 0) {};
	\node [red vertex] (15) at (4, 0) {};
	\node [green vertex] (16) at (6, 0) {};
	\node [red vertex] (17) at (8, 0) {};
	\node [boundary vertex] (18) at (-2.75, -0.5) {};
	\node [boundary vertex] (19) at (1.25, -0.5) {};
	\node [boundary vertex] (20) at (5.25, -0.5) {};
	\node [green vertex] (21) at (-6, -1) {};
	    \node [green angle] at (21) {$\pi$};
	\node [hadamard vertex] (22) at (-1.25, -1) {};
	\node [hadamard vertex] (23) at (2.75, -1) {};
	\node [hadamard vertex] (24) at (6.75, -1) {};
	\node [boundary vertex] (25) at (-2.25, -1.75) {};
	\node [boundary vertex] (26) at (1.75, -1.75) {};
	\node [boundary vertex] (27) at (5.75, -1.75) {};
	\node [boundary vertex] (28) at (-6.75, -2) {};
	\node [green vertex] (29) at (-4, -3) {};
	\node [green vertex] (30) at (0, -3) {};
	\node [green vertex] (31) at (4, -3) {};
	\node [green vertex] (32) at (8, -3) {};
	\node [boundary vertex] (33) at (-4.75, -3.5) {};
	\node [boundary vertex] (34) at (-0.75, -3.5) {};
	\node [boundary vertex] (35) at (3.25, -3.5) {};
	\node [boundary vertex] (36) at (7.25, -3.5) {};
	\draw  (23) to (15);
	\draw  (25) to (22);
	\draw  (21) to (28);
	\draw  (16) to (20);
	\draw  (0) to (8);
	\draw  (1) to (13);
	\draw  (12) to (18);
	\draw  (29) to (33);
	\draw  (30) to (34);
	\draw  (3) to (17);
	\draw  (27) to (24);
	\draw  (0) to (4);
	\draw  (24) to (17);
	\draw  (2) to (15);
	\draw  (16) to (15);
	\draw  (11) to (8);
	\draw  (31) to (35);
	\draw  (26) to (23);
	\draw  (14) to (19);
	\draw  (32) to (36);
	\draw  (17) to (32);
	\draw  (8) to (9);
	\draw  (8) to (12);
	\draw  (8) to (29);
	\draw  (17) to (16);
	\draw  (13) to (12);
	\draw  (21) to (11);
	\draw  (9) to (10);
	\draw  (15) to (14);
	\draw  (22) to (13);
	\draw  (1) to (5);
	\draw  (9) to (17);
	\draw  (15) to (31);
	\draw  (14) to (13);
	\draw  (2) to (6);
	\draw  (3) to (7);
	\draw  (13) to (30);	
\end{tikzpicture}
\quad \quad
\rTo
\quad \quad
\begin{tikzpicture}[quanto]
	\node [boundary vertex] (0) at (-1.5, 4) {};
	\node [boundary vertex] (1) at (-0.5, 4) {};
	\node [boundary vertex] (2) at (0.5, 4) {};
	\node [boundary vertex] (3) at (1.5, 4) {};
	\node [red vertex] (4) at (0, 2.5) {};
	\node [green vertex] (5) at (0, 1) {};
	\node [red vertex] (6) at (0, 1) {};
	\node [red vertex] (7) at (0, -0.5) {};
	\node [boundary vertex] (8) at (-1.5, -2) {};
	\node [boundary vertex] (9) at (-0.5, -2) {};
	\node [boundary vertex] (10) at (0.5, -2) {};
	\node [boundary vertex] (11) at (1.5, -2) {};
	\draw  (7) to (8);
	\draw  (7) to (5);
	\draw  (5) to (4);
	\draw  (2) to (4);
	\draw  (4) to (0);
	\draw  (4) to (1);
	\draw  (4) to (3);
	\draw  (7) to (9);
	\draw  (7) to (11);
	\draw  (10) to (7);
	\node [red angle] at (6) {$\pi$};
\end{tikzpicture}
\]
\end{minipage}}
\caption{A chain of physical $Z$ operations becomes the logical $X_L$ operator.}
\label{chain}
\end{figure}

\indent Within the topological cluster model, the stabilizer operations $X$ and $Z$ on the logical qubit are implemented as physical $Z$ operations (``green $\pi$'') on rings and chains of qubits around or adjacent to defects. A ring is shown in figure \ref{ring}. On the particular colouration that we had for the original cell, this becomes the logical operator $Z_L$ on the logical qubit. A chain of $Z$ operators is shown in figure \ref{chain}. In the cell itself, what changes is the phase of one of the centre red nodes. This then gives the logical operator $X_L$ on the qubit by straightforward application of the rewrites. \\
\indent If we choose a cell on the conjugate lattice, then all colours are reversed. Matching the use in \cite{austin3D}, the red qubit lines as shown in figures \ref{spider}-\ref{chain} represent logical qubits on the dual lattice. Green qubit lines would represent logical qubits on the primal lattice. In that case, it is straightforward to verify that $Z_L$ is constructed from $Z$ chains, and $X_L$ from rings.\\
\indent We can therefore view the primal lattice as supporting green logical nodes, and the dual as supporting red nodes. This highlights an important part of the topological model: there is no straightforward logical Hadamard operator. From the diagrams we can see that what a Hadamard operator would mean is a movement of the qubit from primal to dual lattice or vice versa. This is indeed what the very complex Hadamard procedure shown in \cite{austin2D} performs. When using the red/green calculus for cluster state topological QC, we must bear this in mind: there is no logical H-box in the implementation, only the red and green nodes.

\begin{figure}[t]
\centering
   \scalebox{1}{\input{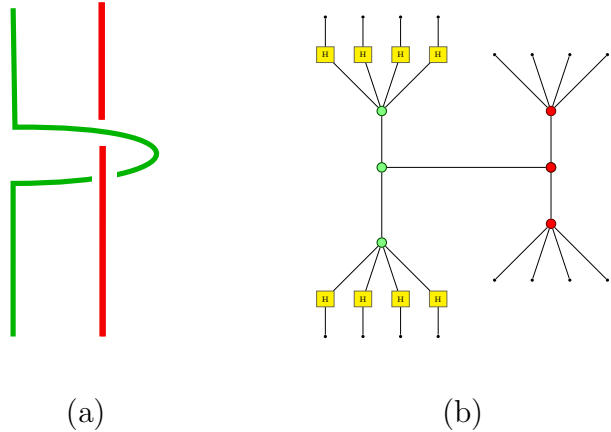}}
\caption{A CNOT braid (a) in the standard representation for topological cluster-state computing, a primal (green) qubit braiding a dual (red) qubit and (b) proposed translation into a red/green pattern (compare with figure \ref{gates}(a)).}
\label{Lcnot}
\end{figure}

\indent One place where this is seen straightforwardly is in the CNOT operation. In the topological model this is performed when the correlation surface from a qubit on the primal lattice (control) intersects with the correlation surface of a dual lattice qubit (target). Writing that using the primal and dual qubit diagrams leads us to propose the diagram for the logical CNOT operation shown in figure \ref{Lcnot}. Compare this now with figure \ref{gates}(a): this is indeed the CNOT gate in the red/green calculus.

\subsection{Double defects}

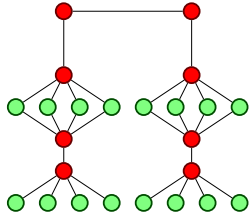
\begin{figure}[t]
\centering
\scalebox{0.85}{ \begin{minipage}[c]{1.0\linewidth}
  \centering

  \[
\begin{tikzpicture}[quanto]
	\node [red vertex] (0) at (-2, 4) {};
	\node [red vertex] (1) at (2, 4) {};
	\node [red vertex] (2) at (-2, 2) {};
	\node [red vertex] (3) at (2, 2) {};
	\node [green vertex] (4) at (-3.5, 1) {};
	\node [green vertex] (5) at (-2.5, 1) {};
	\node [green vertex] (6) at (-1.5, 1) {};
	\node [green vertex] (7) at (-0.5, 1) {};
	\node [green vertex] (8) at (0.5, 1) {};
	\node [green vertex] (9) at (1.5, 1) {};
	\node [green vertex] (10) at (2.5, 1) {};
	\node [green vertex] (11) at (3.5, 1) {};
	\node [red vertex] (12) at (-2, 0) {};
	\node [red vertex] (13) at (2, 0) {};
	\node [red vertex] (14) at (-2, -1) {};
	\node [red vertex] (15) at (2, -1) {};
	\node [green vertex] (16) at (-3.5, -2) {};
	\node [green vertex] (17) at (-2.5, -2) {};
	\node [green vertex] (18) at (-1.5, -2) {};
	\node [green vertex] (19) at (-0.5, -2) {};
	\node [green vertex] (20) at (0.5, -2) {};
	\node [green vertex] (21) at (1.5, -2) {};
	\node [green vertex] (22) at (2.5, -2) {};
	\node [green vertex] (23) at (3.5, -2) {};
	\draw  (14) to (18);
	\draw  (12) to (6);
	\draw  (8) to (13);
	\draw  (3) to (10);
	\draw  (15) to (23);
	\draw  (14) to (17);
	\draw  (15) to (20);
	\draw  (14) to (16);
	\draw  (12) to (4);
	\draw  (12) to (14);
	\draw  (21) to (15);
	\draw  (13) to (11);
	\draw  (5) to (2);
	\draw  (2) to (6);
	\draw  (3) to (9);
	\draw  (15) to (22);
	\draw  (13) to (10);
	\draw  (0) to (2);
	\draw  (4) to (2);
	\draw  (11) to (3);
	\draw  (13) to (15);
	\draw  (9) to (13);
	\draw  (12) to (7);
	\draw  (14) to (19);
	\draw  (5) to (12);
	\draw  (8) to (3);
	\draw  (1) to (3);
	\draw  (7) to (2);
	\draw  (0) to (1);\end{tikzpicture}
\]
\end{minipage}}
\caption{A double defect encoding a single logical qubit line.}
\label{doubd}
\end{figure}
So far we have concentrated on logical qubits created using a single defect line. This has the disadvantage as a topological implementation that logical operators then have to be defined with respect to the external boundaries of the lattice. The standard method of avoiding this is to define logical qubits as pairs of defects, creating in the case of a pair of primal defects the logical entangled state $\alpha \ket{00} + \beta \ket{11}$, which is in itself taken as an encoding of the logical state  $\alpha \ket{0} + \beta \ket{1}$. Logical operators then become either rings of physical $Z$ operators around one of the defects, or chains of operators between the two defects \cite{raussendorf}.

A double defect logical qubit is shown in the diagram of figure \ref{doubd}. In the usual fashion, this has been created from a single large defect that is then split. Simple applications of the spider rule then reduce this to the single logical flow line of the single defect, as in figure \ref{spider}. Showing the logical operators in such a system is not as straightforward, however. The logical $Z$ operation acts only on one of the defects, but the logical $X$ requires operations that touch the pair. We can see how this distinction arises in the qubit calulus, figures \ref{logzz} and \ref{logxx}. Creating the logical $X_L$ from a chain of operators is straightforward as the chain changes both defects identically. The ring of operators around a single defect affects both defects the second defect through the `copying' action of the red nodes on a ``green $\pi$" operator, enabling the $Z$ operator to propagate to the second defect.

\begin{figure}[t]
\centering
\hspace{0cm} \scalebox{0.65}{\hspace{-10cm} \begin{minipage}[c]{1.0\linewidth}
  \centering

  \[
\begin{tikzpicture}[quanto]
	\node [red vertex] (0) at (-2, 4) {};
	\node [red vertex] (1) at (2, 4) {};
	\node [red vertex] (2) at (-2, 2) {};
	\node [red vertex] (3) at (2, 2) {};
	\node [green vertex] (4) at (-3.5, 1) {};
	
	\node [green vertex] (5) at (-2.5, 1) {};
	
	\node [green vertex] (6) at (-1.5, 1) {};
	
	\node [green vertex] (7) at (-0.5, 1) {};
	
	\node [green vertex] (8) at (0.5, 1) {};
	\node [green vertex] (9) at (1.5, 1) {};
	\node [green vertex] (10) at (2.5, 1) {};
	\node [green vertex] (11) at (3.5, 1) {};
	\node [red vertex] (12) at (-2, 0) {};
	\node [red vertex] (13) at (2, 0) {};
	\node [red vertex] (14) at (-2, -2) {};
	\node [red vertex] (15) at (2, -2) {};
	\node [green vertex] (16) at (-3.5, -3) {};
	\node [green vertex] (17) at (-2.5, -3) {};
	\node [green vertex] (18) at (-1.5, -3) {};
	\node [green vertex] (19) at (-0.5, -3) {};
	\node [green vertex] (20) at (0.5, -3) {};
	\node [green vertex] (21) at (1.5, -3) {};
	\node [green vertex] (22) at (2.5, -3) {};
	\node [green vertex] (23) at (3.5, -3) {};
	\draw  (14) to (18);
	\draw  (8) to (13);
	\draw  (3) to (10);
	\draw  (15) to (23);
	\draw  (2) to (7);
	\draw  (14) to (17);
	\draw  (15) to (20);
	\draw  (14) to (16);
	\draw  (2) to (5);
	\draw  (12) to (14);
	\draw  (21) to (15);
	\draw  (13) to (11);
	\draw  (4) to (2);
	\draw  (2) to (6);
	\draw  (3) to (9);
	\draw  (15) to (22);
	\draw  (13) to (10);
	\draw  (0) to (2);
	\draw  (7) to (12);
	\draw  (6) to (12);
	\draw  (11) to (3);
	\draw  (13) to (15);
	\draw  (9) to (13);
	\draw  (5) to (12);
	\draw  (4) to (12);
	\draw  (14) to (19);
	\draw  (8) to (3);
	\draw  (1) to (3);
	\draw  (0) to (1);
	\node [green angle] at (7) {$\pi$};
	\node [green angle] at (5) {$\pi$};
	\node [green angle] at (6) {$\pi$};
	\node [green angle] at (4) {$\pi$};
\end{tikzpicture}
\quad \quad
\rTo
\quad \quad
\begin{tikzpicture}[quanto]
	\node [red vertex] (0) at (-2, 4) {};
	\node [red vertex] (1) at (2, 4) {};
	\node [green vertex] (2) at (-2, 3) {};
	\node [green angle] at (2) {$\pi$};
	\node [red vertex] (3) at (-2, 2) {};
	\node [red vertex] (4) at (2, 2) {};
	\node [green vertex] (5) at (-3.5, 1) {};
	\node [green vertex] (6) at (-2.5, 1) {};
	\node [green vertex] (7) at (-1.5, 1) {};
	\node [green vertex] (8) at (-0.5, 1) {};
	\node [green vertex] (9) at (0.5, 1) {};
	\node [green vertex] (10) at (1.5, 1) {};
	\node [green vertex] (11) at (2.5, 1) {};
	\node [green vertex] (12) at (3.5, 1) {};
	\node [red vertex] (13) at (-2, 0) {};
	\node [red vertex] (14) at (2, 0) {};
	\node [green vertex] (15) at (-2, -1) {};
	\node [green angle] at (15) {$\pi$};
	\node [red vertex] (16) at (-2, -2) {};
	\node [red vertex] (17) at (2, -2) {};
	\node [green vertex] (18) at (-3.5, -3) {};
	\node [green vertex] (19) at (-2.5, -3) {};
	\node [green vertex] (20) at (-1.5, -3) {};
	\node [green vertex] (21) at (-0.5, -3) {};
	\node [green vertex] (22) at (0.5, -3) {};
	\node [green vertex] (23) at (1.5, -3) {};
	\node [green vertex] (24) at (2.5, -3) {};
	\node [green vertex] (25) at (3.5, -3) {};
	\draw  (16) to (20);
	\draw  (3) to (6);
	\draw  (9) to (14);
	\draw  (4) to (11);
	\draw  (17) to (25);
	\draw  (8) to (3);
	\draw  (16) to (19);
	\draw  (17) to (22);
	\draw  (16) to (18);
	\draw  (5) to (3);
	\draw  (13) to (16);
	\draw  (23) to (17);
	\draw  (14) to (12);
	\draw  (4) to (10);
	\draw  (0) to (3);
	\draw  (7) to (13);
	\draw  (17) to (24);
	\draw  (14) to (11);
	\draw  (12) to (4);
	\draw  (14) to (17);
	\draw  (10) to (14);
	\draw  (3) to (7);
	\draw  (6) to (13);
	\draw  (5) to (13);
	\draw  (16) to (21);
	\draw  (1) to (4);
	\draw  (9) to (4);
	\draw  (0) to (1);
	\draw  (8) to (13);
\end{tikzpicture}
\quad \quad
\rTo
\quad \quad
\begin{tikzpicture}[quanto]
	\node [red vertex] (0) at (-2, 4) {};
	\node [red vertex] (1) at (2, 4) {};
	\node [green vertex] (2) at (2, 3) {};
	\node [green angle] at (2) {$\pi$};
	\node [red vertex] (3) at (-2, 2) {};
	\node [red vertex] (4) at (2, 2) {};
	\node [green vertex] (5) at (-3.5, 1) {};
	\node [green vertex] (6) at (-2.5, 1) {};
	\node [green vertex] (7) at (-1.5, 1) {};
	\node [green vertex] (8) at (-0.5, 1) {};
	\node [green vertex] (9) at (0.5, 1) {};
	\node [green vertex] (10) at (1.5, 1) {};
	\node [green vertex] (11) at (2.5, 1) {};
	\node [green vertex] (12) at (3.5, 1) {};
	\node [red vertex] (13) at (-2, 0) {};
	\node [red vertex] (14) at (2, 0) {};
	\node [green vertex] (15) at (-2, -1) {};
	\node [green angle] at (15) {$\pi$};
	\node [red vertex] (16) at (-2, -2) {};
	\node [red vertex] (17) at (2, -2) {};
	\node [green vertex] (18) at (-3.5, -3) {};
	\node [green vertex] (19) at (-2.5, -3) {};
	\node [green vertex] (20) at (-1.5, -3) {};
	\node [green vertex] (21) at (-0.5, -3) {};
	\node [green vertex] (22) at (0.5, -3) {};
	\node [green vertex] (23) at (1.5, -3) {};
	\node [green vertex] (24) at (2.5, -3) {};
	\node [green vertex] (25) at (3.5, -3) {};
	\draw  (16) to (20);
	\draw  (3) to (6);
	\draw  (9) to (14);
	\draw  (4) to (11);
	\draw  (17) to (25);
	\draw  (8) to (3);
	\draw  (16) to (19);
	\draw  (17) to (22);
	\draw  (16) to (18);
	\draw  (5) to (3);
	\draw  (13) to (16);
	\draw  (23) to (17);
	\draw  (14) to (12);
	\draw  (4) to (10);
	\draw  (0) to (3);
	\draw  (7) to (13);
	\draw  (17) to (24);
	\draw  (14) to (11);
	\draw  (12) to (4);
	\draw  (14) to (17);
	\draw  (10) to (14);
	\draw  (3) to (7);
	\draw  (6) to (13);
	\draw  (5) to (13);
	\draw  (16) to (21);
	\draw  (1) to (4);
	\draw  (9) to (4);
	\draw  (0) to (1);
	\draw  (8) to (13);
\end{tikzpicture}
\quad \quad
\rTo
\quad \quad
\begin{tikzpicture}[quanto]
	\node [boundary vertex] (0) at (-1.5, 4) {};
	\node [boundary vertex] (1) at (-0.5, 4) {};
	\node [boundary vertex] (2) at (0.5, 4) {};
	\node [boundary vertex] (3) at (1.5, 4) {};
	\node [red vertex] (4) at (0, 2.5) {};
	\node [green vertex] (5) at (0, 1) {};
	\node [green vertex] (6) at (0, 1) {};
	\node [red vertex] (7) at (0, -0.5) {};
	\node [boundary vertex] (8) at (-1.5, -2) {};
	\node [boundary vertex] (9) at (-0.5, -2) {};
	\node [boundary vertex] (10) at (0.5, -2) {};
	\node [boundary vertex] (11) at (1.5, -2) {};
	\draw  (7) to (8);
	\draw  (7) to (5);
	\draw  (5) to (4);
	\draw  (2) to (4);
	\draw  (4) to (0);
	\draw  (4) to (1);
	\draw  (4) to (3);
	\draw  (7) to (9);
	\draw  (7) to (11);
	\draw  (10) to (7);
	\node [green angle] at (6) {$\pi$};
\end{tikzpicture}
\]
\end{minipage}}
\caption{A ring of phyical $Z$ operators on one defect becomes the logical $Z_L$ operator.}
\label{logzz}
\end{figure}
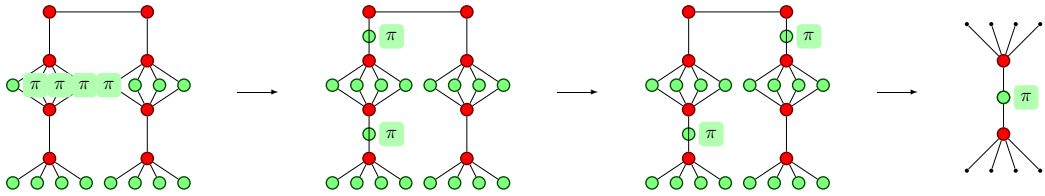
\begin{figure}[t]
\centering
\hspace{0cm} \scalebox{0.9}{\begin{minipage}[c]{1.0\linewidth}
  \centering

  \[
\begin{tikzpicture}[quanto]
	\node [red vertex] (0) at (-2, 4) {};
	\node [red vertex] (1) at (2, 4) {};
	\node [red vertex] (2) at (-2, 2) {};
	\node [red vertex] (3) at (2, 2) {};
	\node [green vertex] (4) at (-3.5, 1) {};
	\node [green vertex] (5) at (-2.5, 1) {};
	\node [green vertex] (6) at (-1.5, 1) {};
	\node [green vertex] (7) at (-0.5, 1) {};
	\node [green vertex] (8) at (0.5, 1) {};
	\node [green vertex] (9) at (1.5, 1) {};
	\node [green vertex] (10) at (2.5, 1) {};
	\node [green vertex] (11) at (3.5, 1) {};
	\node [red vertex] (12) at (-2, 0) {};
	\node [red vertex] (13) at (2, 0) {};
	\node [red vertex] (14) at (-2, -1) {};
	\node [red angle] at (14) {$\pi$};
	\node [red vertex] (15) at (2, -1) {};
	\node [red angle] at (15) {$\pi$};
	\node [red vertex] (16) at (-2, -2) {};
	\node [red vertex] (17) at (2, -2) {};
	\node [green vertex] (18) at (-3.5, -3) {};
	\node [green vertex] (19) at (-2.5, -3) {};
	\node [green vertex] (20) at (-1.5, -3) {};
	\node [green vertex] (21) at (-0.5, -3) {};
	\node [green vertex] (22) at (0.5, -3) {};
	\node [green vertex] (23) at (1.5, -3) {};
	\node [green vertex] (24) at (2.5, -3) {};
	\node [green vertex] (25) at (3.5, -3) {};
	\draw  (16) to (20);
	\draw  (12) to (6);
	\draw  (8) to (13);
	\draw  (3) to (10);
	\draw  (17) to (25);
	\draw  (16) to (19);
	\draw  (17) to (22);
	\draw  (16) to (18);
	\draw  (12) to (4);
	\draw  (12) to (16);
	\draw  (23) to (17);
	\draw  (13) to (11);
	\draw  (5) to (2);
	\draw  (2) to (6);
	\draw  (3) to (9);
	\draw  (17) to (24);
	\draw  (13) to (10);
	\draw  (0) to (2);
	\draw  (4) to (2);
	\draw  (11) to (3);
	\draw  (13) to (17);
	\draw  (9) to (13);
	\draw  (12) to (7);
	\draw  (16) to (21);
	\draw  (5) to (12);
	\draw  (8) to (3);
	\draw  (1) to (3);
	\draw  (7) to (2);
	\draw  (0) to (1);
\end{tikzpicture}
\quad \quad
\rTo
\quad \quad
\begin{tikzpicture}[quanto]
	\node [boundary vertex] (0) at (-1.5, 4) {};
	\node [boundary vertex] (1) at (-0.5, 4) {};
	\node [boundary vertex] (2) at (0.5, 4) {};
	\node [boundary vertex] (3) at (1.5, 4) {};
	\node [red vertex] (4) at (0, 2.5) {};
	\node [green vertex] (5) at (0, 1) {};
	\node [red vertex] (6) at (0, 1) {};
	\node [red vertex] (7) at (0, -0.5) {};
	\node [boundary vertex] (8) at (-1.5, -2) {};
	\node [boundary vertex] (9) at (-0.5, -2) {};
	\node [boundary vertex] (10) at (0.5, -2) {};
	\node [boundary vertex] (11) at (1.5, -2) {};
	\draw  (7) to (8);
	\draw  (7) to (5);
	\draw  (5) to (4);
	\draw  (2) to (4);
	\draw  (4) to (0);
	\draw  (4) to (1);
	\draw  (4) to (3);
	\draw  (7) to (9);
	\draw  (7) to (11);
	\draw  (10) to (7);
	\node [red angle] at (6) {$\pi$};
\end{tikzpicture}
\]
\end{minipage}}
\caption{A chain of phyical $Z$ operators betwen the two defects becomes the logical $X_L$ operator.}
\label{logxx}
\end{figure}
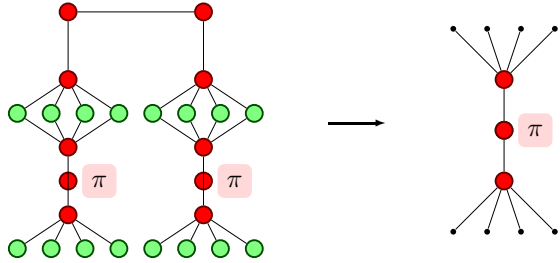

\subsection{Homology}

Homological equivalence plays an important role in topological computing. Equivalent patterns of defects implement the same algorithm, and this method can be used to re-write circuits according to various rules \cite[\S 2]{raussendorf}. This same notion is present in quantum picturalism, described intuitively as ``only the topology matters" \cite{monster}: continuously deforming diagrams does not change the processes represented. It is not, however, immediately obvious that we can derive this from the red/green calculus alone for topological QC. There are two main steps in using the pictorial representation: first we re-write the diagrams of individual physical qubits to give logical qubit diagrams, and then those logical qubit diagrams themselves can be re-written into others. If we are to change the geometrical structure of the defects to a homologically identical one, then this will in general change the relative positions of the qubits that are to be measured out to form defects. In particular, they will in general take up different relative positions in the $z$-dimension (depth) of the cluster in homologically equivalent defect patterns -- that is, they will be measured out at different relative computational times. Recall that we can only begin to re-write the physical qubit diagram once the measurement has happened; this means that the order in which the rewrite rules are applied to the nodes making up the logical qubit line is changed.

In general, the red/green calculus and related systems are not insensitive to the order in which rewrite rules are applied. The relevant property of the system is called \emph{confluence}: for all sequences of the set of reduction relations (rewrites) in a language, if any permutation of a given rewrite sequence yields the same output form as any other, then the language is confluent. It is known that, in general, the red/green calculus is not confluent \cite{confluence}. However, it is also known that confluence in the calculus is broken only by a single rule, the \emph{bialgebra} re-write rule. Without the bialgebra rule the red/green system is confluent \cite{confluence}. The scheme that we've presented does not rely on the bialgebra rule to rewrite physical to logical qubit diagrams; these re-writing results are therefore unchanged under a different order of application of the re-writes, and therefore do not change when the physical geometry of the defect patterns are changed.

This leaves us with an important restriction on the use of the red/green calculus in topological computing. At no stage can the bialgebra rule be used for rewriting, either for diagrams of physical qubit processes or when re-writing logical qubit lines.

\section{Automated design and reasoning}

One particular consequence of the use of the red/green calculus to describe topological QC is that process re-writing is available at two separate levels. Re-writing is an important tool as it enables us to give different patterns of measurements in the cluster that implement the same algorithm, and also because it allows us to demonstrate the algorithm implemented by a given -- potentially highly complex -- sequence of defect patterns. 

 The first type, a re-writing of the logical qubit diagrams, allows us to define multiple physically implementable measurement patterns for a given algorithms. These can then be used in optimisation procedures, given a particular set of constraints. There has been this ability in the topological model previously, given by the homological re-write rules of \cite{raussendorf}. The red/green re-writes can supplement topologically simple re-writes, allowing for circuit identities that are not necessarily continuous deformations of each other. This greater variety of applicable re-writes is particularly important for circuit optimisation problems, which will in general be performing some form of search over a set of potential circuit implementations.
 
 The second form of re-writing, that of finding the equivalent logical pattern from a set of measurements on individual qubits, gives a powerful method for verifying the algorithm implemented without calculating and manipulating stabilizers or other state descriptors for each separate physical qubit. Even for small lattice sizes working through sets of stabilizers can become technically prohibitive -- even for a relatively small code distance, the cross-section of the lattice can be of the order of a a few hundred qubits square, and while we can easily lay out standard measurement patterns for an algorithm on that lattice, verifying the result of an arbitrary measurement pattern through stabilizer operations is unfeasible.
 
 For these larger numbers of qubits, of course, even the rewriting would be impractical to perform by hand. However, the red/green calculus lends itself very naturally to automation, given the particular set of re-write rules. A tool for automatic re-writes has already been developed \cite{quantomatic}, giving a first hint of the possibilities available for automatic design and verification tools for quantum algorithms were similar tools to be developed for individual physical implementations. Such tools may in the future play a vital role in compiling quantum programs to run on specific physical systems under their individual design constraints.

\section{Conclusions}

In this paper we demonstrate the use of the diagrammatic representation of quantum picturalism as a high-level language to describe the patterns of excitation in topological cluster-state quantum computing. We have shown that qubit lines in the category-theory based red/green calculus are geometrically identical with the physical defect lines within the cluster used to implement a given algorithm. In doing this, we have demonstrated the structure of large-scale cluster states within the calculus, and given a representation for stabilizers. Using the pictorial language has enabled us to give intuitively straightforward descriptions of elements in the topological model such as the primal and dual lattice, logical qubits, operators and homology, and brings out aspects of the information flow within the model that are opaque in standard treatments. 

The structural equivalence between the defect patterns and category-theory diagrams is somewhat surprising. The diagrammatic language represents processes rather than qubits as vertices in its graphs. What the equivalence tells us is that it is fundamentally these processes that are represented in topological cluster-state QC: it is not, structurally, a model of single qubits evolving over time. Instead, such a situation is simulated using many entangled qubits (or qubit regions) to represent processes on an abstract ``logical qubit". Entanglement between these qubits simulates the persistence and evolution of a single qubit over time. At a fundamental level, topological cluster-state QC is described by a process algebra rather than a qubit algebra.

As well as such fundamental implications, the use of quantum picturalism for topological QC opens several avenues for the design and verification of quantum algorithms on cluster states, and the ability to automate these procedures (vital when large numbers of physical qubits are involved). Diagrams can be translated directly to patterns of measurement on individual qubits, and \emph{vice versa}, making the red/green calculus a true `compiler language' for topological QC: as well as enabling high-level reasoning, it also allows us to implement the high-level concept directly in the physical structure of the cluster state. The circuit and pattern rewriting capabilities of the red/green calculus lend themselves naturally to automation, one tool for which already exists within the community. 

%
This work opens up new possibilities for both the topological model and the category-theoretic framework. The fact that the defects in the 3D topological model can be seen as a physical instantiation of the logical flows in category-theory based diagrammatics gives a direct physical model and application for quantum picturalism. Conversely, we now have a language that is both graphical and native to the topological model to describe computations, without needing to transform back to the circuit model. Quantum picturalism may in fact end up as the first deployed quantum programming language.

\subsubsection*{Acknowledgements}

I gratefully acknowledge many interesting and useful discussions and pub sessions with the Oxford quantum category theorists. In particular I would like to thank Aleks Kissenger, Ross Duncan, and Bob Coecke for discussions and comments on the paper. Thanks also to Robert Raussendorf for pointing out an error in a previous version, and to Rodney Van Meter for close reading of the manuscript. I acknowledge support from EU project QICS. This research is supported by the JSPS through its FIRST Program.\\







%
%


\begin{thebibliography}{000}

\bibitem{shorec} P.W. Shor, `Fault-tolerant quantum computation,' \emph{FOCS}, pp.56, 37th Annual Symposium on Foundations of Computer Science (FOCS '96) (1996).
\bibitem{kitaev} A. Y. Kitaev, `Fault-tolerant quantum computing by anyons', \emph{Annals of Physics} 303 2-30 (2003).
\bibitem{monster} B. Coecke and R. Duncan, `Interacting Quantum Observables: Categorical Algebra and Diagrammatics', \emph{New J. Physics} 13 043016 (2011).
\bibitem{bobsamson1} S. Abramsky and B. Coecke, `A categorical semantics of quantum protocols', in \emph{Proceedings of the 19th IEEE conference on Logic in Computer Science} (LiCS'04). IEEE Computer Science Press (2004).
\bibitem{bob1} B. Coecke, `Quantum picturalism', \emph{Contemporary Physics} 51, 59-83 (2010).
\bibitem{penrose} R. Penrose, `Applications of negative dimensional tensors', in \emph{Combinatorial Mathematics and its Applications} 221-244. Academic Press. (1971).
\bibitem{joyal} A. Joyal and R. Street, `The geometry of tensor calculus', \emph{Advances in Mathematics} 88, 55-112 (1991).
\bibitem{selinger} P. Selinger, `Dagger compact closed categories and completely positive maps', \emph{ENTCS} 170, 139-163 (2005).
\bibitem{anyon} M. H. Freedman, A. Kitaev, M. J. Larsen, Z. Wang, `Topological Quantum Computation', arXiv:quant-ph/0101025 (2001).
\bibitem{jiannis} G. K. Brennen and J. K. Pachos, `Why should anyone care about computing with anyons?', \emph{Proceedings of the Royal Society} A 464, 1-24 (2008).
\bibitem{bravyikitaev} S. B. Bravyi and A. Y. Kitaev, `Quantum codes on a lattice with boundary', 
\bibitem{surface} R. Raussendorf and J. Harrington, `Fault-Tolerant Quantum Computation with High Threshold in Two Dimensions' \emph{Phys. Rev. Lett.} 98, 190504 (2007).
\bibitem{MBQC} R. Raussendorf, D. Browne and H.-J. Briegel `Measurement-based quantum computation on cluster states', \emph{Physical Review A} 68, 022312 (2003).
\bibitem{ross} R. Duncan and S. Perdrix, `Graph states and the necessity of Euler decomposition', Mathematical Theory and Computational Practice
Lecture Notes in Computer Science Volume 5635 (2009).
\bibitem{rosssimon} R. Duncan and S. Perdrix, `Rewriting Measurement-Based Quantum Computations with Generalised Flow' in `Automata, Languages and Programming' \emph{Lecture Notes in Computer Science} 6199/2010, 285-296 (2010).
\bibitem{raussendorf} R. Raussendorf, J. Harrington and K. Goyal, `Topological fault-tolerance in cluster state quantum computation' \emph{New J. Phys} 9, 199 (2007).
\bibitem{rg} B. Coecke and R. Duncan, `Interacting quantum observables', in L. Aceto \emph{et al} (eds.) ICALP 2008, Part II. \emph{LNCS} vol. 5126, pp. 298-310 (2008).
\bibitem{austin2D}A. G. Fowler, A. M. Stephens, and P.Groszkowski, `High-threshold universal quantum computation on the surface code' \emph{Phys. Rev. A} 80, 052312 (2009).
\bibitem{austin3D} A. G. Fowler and K. Goyal, `Topological cluster state quantum computing' \emph{Quant. Info. Comput.} 9, 721-738 (2009).
\bibitem{onepc} David S. Wang, Austin G. Fowler and Lloyd C. L. Hollenberg, `Quantum computing with nearest neighbor interactions and error rates over 1\%', \emph{Phys. Rev. A} 83, 020302 (2011).
\bibitem{simon} Simon J. Devitt, Austin G. Fowler, Ashley M. Stephens, Andrew D. Greentree, Lloyd C.L. Hollenberg, William J. Munro and Kae Nemoto, `Architectural design for a topological cluster state quantum computer', \emph{New. J. Phys.} 11, 083032 (2009). 
\bibitem{rod} Rodney Van Meter, Thaddeus D. Ladd, Austin G. Fowler and Yoshihisa Yamamoto, `Distributed quantum Ccomputation architecture using semiconductor nanophotonics', \emph{Int. J. Quantum Inf.} 8 295-323 (2010). 
 \bibitem{exec}Philipp Schindler, Julio T. Barreiro, Thomas Monz, Volckmar Nebendahl, Daniel Nigg, Michael Chwalla, Markus Hennrich and Rainer Blatt, `Experimental Repetitive Quantum Error Correction', \emph{Science} 332, 1059 (2011).
\bibitem{gottesman} D. Gottesman, PhD thesis, Caltech. quant-ph/9705052 (1997).
\bibitem{confluence} A. Kissinger, Masters thesis, Oxford. http://www.comlab.ox.ac.uk/people/aleks.kissinger/papers.html [accessed 21/01/2011](2008).
\bibitem{quantomatic} L. Dixon, R. Duncan, A. Kissinger, and A. Merry, \emph{Quantomatic} http://dream.inf.ed.ac.uk/projects/quantomatic/
\end{thebibliography}
\end{document}